%% file: main.tex
\documentclass[sigplan,screen,prologue,dvipsnames]{acmart}
\AtBeginDocument{%
  \providecommand\BibTeX{{%
    \normalfont B\kern-0.5em{\scshape i\kern-0.25em b}\kern-0.8em\TeX}}}

\copyrightyear{2020}
\acmYear{2020}
\setcopyright{acmlicensed}
\acmConference[ASPLOS '20]{Proceedings of the Twenty-Fifth International Conference on Architectural Support for Programming Languages and Operating Systems}{March 16--20, 2020}{Lausanne, Switzerland}
\acmBooktitle{Proceedings of the Twenty-Fifth International Conference on Architectural Support for Programming Languages and Operating Systems (ASPLOS '20), March 16--20, 2020, Lausanne, Switzerland}
\acmPrice{15.00}
\acmDOI{10.1145/3373376.3378514}
\acmISBN{978-1-4503-7102-5/20/03}

\author{Xuan Yang}
\email{xuany@stanford.edu}
\affiliation{%
  \institution{Stanford University}
}

\author{Mingyu Gao}
\email{gaomy@tsinghua.edu.cn}
\affiliation{%
  \institution{Tsinghua University}
}

\author{Qiaoyi Liu}
\email{joeyliu@stanford.edu}
\affiliation{%
  \institution{Stanford University}
}

\author{Jeff Setter}
\email{setter@stanford.edu}
\affiliation{%
  \institution{Stanford University}
}

\author{Jing Pu}
\email{jingpu@alumni.stanford.edu}
\affiliation{%
  \institution{Stanford University}
}

\author{Ankita Nayak}
\email{ankitan@stanford.edu}
\affiliation{%
  \institution{Stanford University}
}

\author{Steven Bell}
\email{sebell@stanford.edu}
\affiliation{%
  \institution{Stanford University}
}

\author{Kaidi Cao}
\email{kaidicao@stanford.edu}
\affiliation{%
  \institution{Stanford University}
}

\author{Heonjae Ha}
\email{hunjaeha@stanford.edu}
\affiliation{%
  \institution{Stanford University}
}

\author{Priyanka Raina}
\email{praina@stanford.edu}
\affiliation{%
  \institution{Stanford University}
}

\author{Christos Kozyrakis}
\email{kozyraki@stanford.edu}
\affiliation{%
  \institution{Stanford University, Google}
}

\author{Mark Horowitz}
\email{horowitz@ee.stanford.edu}
\affiliation{%
  \institution{Stanford University}
}

\renewcommand{\shortauthors}{Yang et al.}



\usepackage[normalem]{ulem}

\usepackage{graphicx}
\usepackage{subfig}
\usepackage{booktabs} 

\usepackage{algorithm}
\usepackage[noend]{algorithmic}

\usepackage{amsmath}
\usepackage{listings}
\usepackage[flushleft]{threeparttable}
\usepackage{multirow}
\usepackage{comment}
\usepackage{float}
\usepackage{enumitem}

\definecolor{codeblue}{rgb}{0.2,0.2,1}
\definecolor{codegreen}{rgb}{0,0.6,0}
\definecolor{codegray}{rgb}{0.5,0.5,0.5}
\definecolor{codepurple}{rgb}{0.58,0,0.82}
\definecolor{backcolour}{rgb}{0.95,0.95,0.92}
\lstdefinestyle{mystyle}{
    commentstyle=\color{codegreen},
    keywordstyle=\color{codeblue}\bfseries,
    numberstyle=\tiny\color{codegray},
    stringstyle=\color{codepurple},
    basicstyle=\footnotesize\ttfamily,
    breakatwhitespace=false,         
    breaklines=true,                 
    captionpos=b,                    
    keepspaces=true,                 
    numbers=left,                    
    numbersep=5pt,                  
    showspaces=false,                
    showstringspaces=false,
    showtabs=false,
    frame=single,
    tabsize=4,
    columns=flexible
}
\begin{document}
\fancyhead{}
\title{
Interstellar: Using Halide's Scheduling Language to 
Analyze DNN Accelerators}

\begin{abstract}
\input{abstract}
\end{abstract}

\date{}
\maketitle

\thispagestyle{empty}

\input{intro}
\input{background}
\input{taxonomy}

\input{implementation}

\input{method}

\input{result}
\input{conclusion}
\input{ack}


\bibliographystyle{plain}
\balance
\bibliography{ref}

\end{document}

%% file: abstract.tex
We show  that DNN accelerator micro-architectures and their program mappings represent specific choices of loop order and hardware parallelism for computing  the seven nested loops of DNNs, which enables us to create a formal taxonomy of all existing dense DNN accelerators. Surprisingly, the loop transformations needed to create these hardware variants can be precisely and concisely represented by Halide's scheduling language. By modifying the Halide compiler to generate hardware, we create a system that can fairly compare these prior accelerators. As long as proper loop blocking schemes are used, and the hardware can support mapping replicated loops, many different hardware dataflows yield similar energy efficiency with good performance. This is because the loop blocking can ensure that most data references stay on-chip with good locality and the processing units have high resource utilization. How resources are allocated, especially in the memory system, has a large impact on energy and performance.  By optimizing hardware resource allocation while keeping throughput constant, we achieve up to 4.2$\times$ energy improvement for Convolutional Neural Networks (CNNs), 1.6$\times$ and 1.8$\times$ improvement for Long Short-Term Memories (LSTMs) and multi-layer perceptrons (MLPs), respectively.

%% file: intro.tex
\section{Introduction}

Deep neural networks (DNNs) have recently displaced classical image processing and machine learning methods due to their state-of-the-art performance on many tasks, particularly in recognition, localization, and detection~\cite{hinton2006fast}. As the number of applications for DNNs has grown, so have proposals for DNN accelerators. NeuFlow created a 2D systolic array for convolutional neural networks (CNNs)~\cite{farabet2011neuflow}. The DianNao family was built around customized inner-product units~\cite{chen2014diannao, chen2014dadiannao}. Eyeriss highlighted the importance of on-chip dataflow on energy efficiency and proposed a row-stationary heuristic~\cite{eyeriss_isca16, eyeriss_isscc16}. And Google's TPU used a simple yet efficient systolic dataflow on a large 2D processing array~\cite{tpu_isca17}. These are just a few of the recent publications on DNN acceleration. All these proposals stated the advantages of their approaches over conventional general-purpose baseline platforms, yet the architectures and dataflows differ significantly across these approaches.

To help us understand and further improve these DNN accelerators, we realized that, since all the hardware designs perform the same computation, i.e., a seven-level loop nest for convolution as in Algorithm~\ref{alg:sim_alg}, the space of all accelerators can be formally specified by how they transform (block, reorder, and parallelize) the loop nest. We use this insight to create a formal taxonomy of DNN accelerators that expresses design choices as different loop transformations. For example, to improve data reuse and energy efficiency, the loops can be blocked and reordered to better schedule the computation, such that most data references are captured by the smallest and the most efficient memory buffer. The on-chip dataflow choices can also be represented as parallelizing different loops on multiple hardware processing units, known as spatial loop unrolling.


Prior work, like Timeloop~\cite{timeloop}, also adopted similar loop-based approaches to represent and analyze the design space for DNN accelerators systematically. We extend this work by showing that the loop transformations we use to specify the micro-architecture and dataflow choices of DNN accelerators are almost a subset of the loop transformation and memory allocation primitives provided by Halide's scheduling language~\cite{halide2013}. Halide is a domain-specific language for image processing applications. It provides all the required facilities to perform these loop transformations to convert a single application into efficient implementations on CPU, GPU, and more recently specialized hardware~\cite{halidefpga2017}. 

To create any possible DNN dataflow and storage hierarchy, we extend Halide, enabling it to create hardware accelerators for dense linear algebra in addition to image processing. The decoupling between algorithm and schedule within the Halide system makes it easy to explore different DNN mappings and hardware by simply changing the schedules in the Halide code. Using this system makes it easy to recreate the previously proposed designs and fairly compare their resulting performance and energy efficiency.


The extended Halide system allows us to create a systematic optimization framework, which can efficiently study the impact of dataflow and underlying hardware architecture design choices. Our results using Halide and the optimization framework show, with proper loop blocking, many dataflow choices can achieve similar and close-to-optimal energy efficiency. This large number of ``optimal'' solutions results from the fact that operands in most convolutional layers in DNNs have high reuse rates so, as long as properly blocked, most data references occur locally. Tailoring the loop blocking to each architecture is key; the blocking approach is usually more critical than the dataflow choice. For the layers that do not have enough data reuse to exploit, e.g., fully-connected layers that are not batched together, overall energy is dominated by off-chip main memory accesses, thus the on-chip dataflow still does not have a large impact. From a performance perspective, it is important that the hardware supports unrolling multiple loops onto one spatial dimension of the underlying hardware, named as replication, to achieve good resource utilization.

In fact, energy efficiency is more tightly tied to the design of the hierarchical memory system and how each level in this hierarchy is sized. Since every multiply-add (MAC) involves fetching many operands from a memory (usually a register file, RF) and the cost of each RF fetch is proportional to the RF size, it is most efficient to adopt a relatively small RF. This small first-level memory creates the need for a memory hierarchy, since the size ratio between the adjacent memory levels needs to be in a certain range to balance the total energy cost of accessing data at each level in the memory hierarchy.
Using these insights, we created an efficient optimizer for these types of Halide programs to jointly optimize memory system with schedules, which achieves up to 4.2$\times$, 1.6$\times$, and 1.8$\times$ energy improvement over the original Eyeriss accelerator for various CNNs, LSTMs, and MLPs respectively.

This paper makes the following contributions:
\begin{itemize}[topsep=0pt,itemsep=0pt]
\item Introduces a systematic approach to precisely and concisely describe the design space of DNN accelerators as schedules of loop transformations.
\item Shows that both the micro-architectures and dataflow mappings for existing DNN accelerators can be expressed as schedules of a Halide program, and extends the Halide schedule language and the Halide compiler to produce different hardware designs in the space of dense DNN accelerators.
\item Creates a tool to optimize the memory hierarchy, which is more important than the choice of dataflow, achieving a $1.8\times$ to $4.2\times$ energy improvement for CNNs, LSTMs, and MLPs.
\end{itemize}

The next section briefly reviews DNN accelerators. Then Section~\ref{sec:design_space} describes how these accelerators can be characterized by their loop nest structures. Section~\ref{sec:halideHardware} introduces Halide, explains how its scheduling language expresses the transformations we need, and shows how we use it to generate different accelerator implementations. To help us rapidly evaluate these designs, Section~\ref{sec:method} discusses an analytical model, and how we validated this model. We then use this model to evaluate the energy and performance of different designs in Section~\ref{sec:result}. Finally Section~\ref{sec:conclusion} concludes the paper.

%% file: background.tex
\section{Diversity of DNN Accelerators}
\label{sec:background}

While independent research efforts often converge to a few common approaches, this does not seem to be the case for DNN acceleration. The NeuFlow architecture was a 2D systolic array for CNNs, where each processing element (PE) communicated only with its neighbors, and data was streamed to and from DRAM~\cite{farabet2011neuflow}. Its successor TeraDeep used a fixed loop-blocking strategy for CONV layers~\cite{gokhale2014teradeep}. The DianNao family was built around customized inner-product units. The first generation used a single level of small buffers~\cite{chen2014diannao}, while in a later iteration the original unit was surrounded by a large eDRAM that stored the complete data sets~\cite{chen2014dadiannao}. Another version specially built for embedded systems further extended to a 2D PE mesh that supported optimized inter-PE data propagation~\cite{shidiannao_isca15}. More recently, Eyeriss highlighted the importance of such on-chip dataflow for energy efficiency, and proposed using row-stationary dataflow as a heuristic solution~\cite{eyeriss_isca16, eyeriss_isscc16}. 
Neurocube and Tetris combined the spatial PE arrays with 3D-stacked DRAM to reduce the main memory access cost~\cite{tetris_asplos17,  neurocube_isca16}. FlexFlow leveraged the complementary effects of different dataflow styles and mixed them on the same PE array to improve resource utilization~\cite{flexflow_hpca17}. To improve its efficiency, Eyeriss V2~\cite{chen2018eyeriss} designed a flexible interconnect to support different replication schemes. In addition to CNNs, Google's TPU used a simple systolic dataflow on a large 2D array of PEs, which could also be used for MLPs and LSTMs~\cite{tpu_isca17}. Tangram investigates the dataflow optimizations for coarse-grain parallelism to eliminate excessive data
duplication in the on-chip buffers for tiled NN accelerators~\cite{Gao:2019:TOC:3297858.3304014}. Song in~\cite{song2018towards} proposed to reorganize dataflows to improve the data reuse for non-standard convolutional layers in Generative Adversarial Networks (GANs). Other prior work has also implemented architectures that are flexible to support multiple different dataflow types~\cite{Kwon:2018:MEF:3173162.3173176, Wei:2017:ASA:3061639.3062207}. Beyond dense matrix computations, other designs have explored DNN sparsity and proposed specialized dataflow schedules~\cite{eie_isca16, cnvlutin_isca16, cambricon_x_micro16, parashar2017scnn, scalpel_isca17}. Another group of designs have transformed DNN processing into the frequency domain to reduce computations~\cite{Zhang:2017:FDA:3020078.3021727, Ko:2017:DEA:3061639.3062228}.

On FPGA platforms, Zhang et al.~\cite{zhang2015optimizing} adopted the Roofline model to explore loop blocking, but considered only two levels of memory and only minimized off-chip bandwidth rather than total memory energy. Alwani et al.~\cite{fused_layer_cnn_micro16} fused the computation of different NN layers, and Li et al.~\cite{cnn_pipeline_fpl16} mapped the entire CNN onto an FPGA in a pipelined manner, both to reduce intermediate data writeback. Shen et al.~\cite{cnn_fpga_util_fpl16, cnn_fpga_util_isca17} optimized FPGA resource utilization by using a heterogeneous design. Sharma et al.~\cite{7783720} provided hand-optimized templates for generating dataflow architectures. 

%% file: taxonomy.tex
\section{DNN Accelerator Design Space}
\label{sec:design_space}

\begin{algorithm}[tb]
\begin{algorithmic}
    \FOR{$b = 0$ {\bf until} $B$}
    \FOR{$k = 0$ {\bf until} $K$}
    \FOR{$c = 0$ {\bf until} $C$}
    \FOR{$y = 0$ {\bf until} $Y$}
    \FOR{$x = 0$ {\bf until} $X$}
    \FOR{$f_y = 0$ {\bf until} $F_Y$}
    \FOR{$f_x = 0$ {\bf until} $F_X$}
        \STATE $\mathbf{O}[b][k][x][y] \mathrel{+}= \mathbf{I}[b][c][x\!+\!f_x][y\!+\!f_y]$ \\\hspace{6em} $\times \mathbf{W}[k][c][f_x][f_y]$
    \ENDFOR\ENDFOR\ENDFOR\ENDFOR\ENDFOR\ENDFOR\ENDFOR
\end{algorithmic}
\caption{CONV layer: simple seven nested loops.}
\label{alg:sim_alg}
\end{algorithm}

All the DNN accelerators discussed in Section~\ref{sec:background} have demonstrated improvements over general-purpose baselines, and explored how the parameters they experimented with would affect their performance and efficiency.  Unfortunately, without an understanding of the global design space, each paper explores a different part of the space\,---\,perhaps coupling together independent parameters\,---\,so this approach leads to conflicting reports on the ``optimal'' parameters. To avoid this problem we first wanted to understand the space of all possible dense DNN accelerators. We thought this was possible since each accelerator computes the same result, so the differences must be in the resources available to compute the result, and the way the computation is scheduled to use these resources.  To understand how this works, let's review the computation which needs to be performed.


A DNN is a directed acyclic graph (DAG) of various types of layers. The CONV layer computation is summarized as:
\begin{equation*}
\begin{split}
& \mathbf{O}[b][k][x][y] = \\
& \quad \sum_{c=0}^{C-1}{\sum_{f_y=0}^{F_Y-1}{\sum_{f_x=0}^{F_X-1}{
\mathbf{I}[b][c][x\!+\!f_x][y\!+\!f_y] \times \mathbf{W}[k][c][f_x][f_y]}}}
\end{split}
\end{equation*}
and also shown in Algorithm~\ref{alg:sim_alg} as seven levels of nested loops. The nested loops generate output feature maps (\emph{fmaps}) $\mathbf{O}$, which have $K$ channels of $X\!\times\!Y$ images, by processing the input fmaps $\mathbf{I}$ of $C$ channels. The fmap data are processed in \emph{batches}  ($B$) to increase parallelism and data reuse. $\mathbf{W}$ contains the weights as $K$ 3D stencil filters with size $C\!\times\!F_X\!\times\!F_Y$. By summarizing this computation by this loop nest, we can express computation beyond the batched CONV layers, including non-batched operations, fully-connected (FC) layers, etc., by setting some loop bounds to 1. For example, the FC layer computes a matrix vector multiplication and can be described using the same nested loops with only $C$, $K$, and $B$ loops, while all the other loops bounds are set to 1.

DNNs also include other layer types such as pooling, normalization, and element-wise. However, CONV and FC layers dominate the computation and memory communication, so we focus on them in this paper.

\subsection{Design Space Overview}

\begin{figure}
    \centering
    \includegraphics[width=0.45\textwidth]{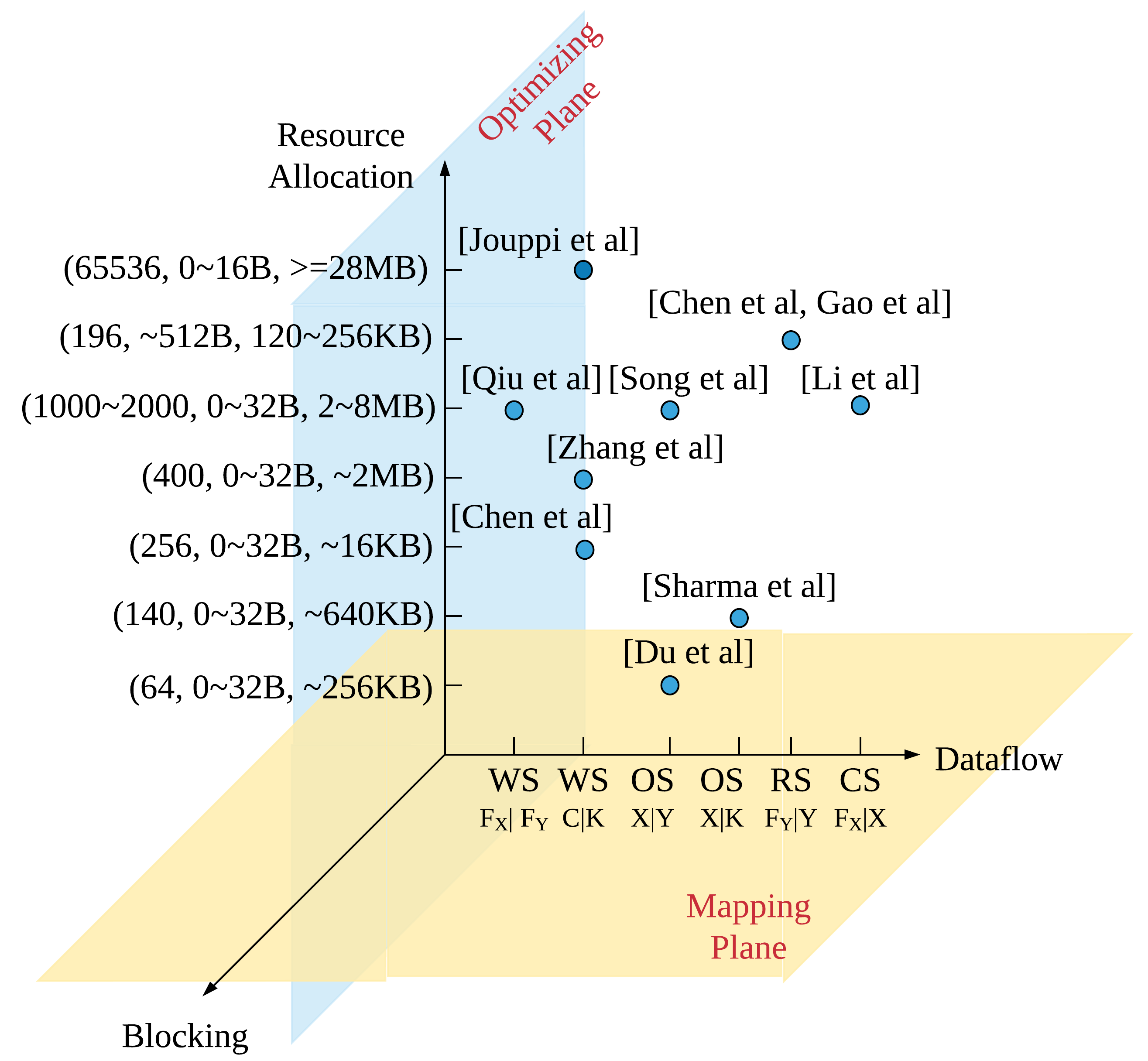}
    \caption{3D design space for DNN accelerators. The positions of labels or vectors on each axis only represent different choices without specific information about ordering or distance.}
    \label{fig:3d_space}
\end{figure}

The design parameters that have been widely studied to optimize data reuse and performance of DNNs are \emph{loop blocking}~\cite{blocking_cnn_arxiv16} and \emph{dataflow}~\cite{eyeriss_isca16, DBLP:journals/corr/abs-1805-02566}. However, when exploring such software scheduling decisions, prior designs often used different hardware \emph{resource allocations}. We therefore consider a corresponding three-dimensional design space for DNN accelerators, as shown in Figure~\ref{fig:3d_space}.

\textbf{Dataflow:} DNN accelerators often exploit parallelism to improve performance by using multiple processing elements (PEs) simultaneously. Essentially it executes one or more loops in Algorithm \ref{alg:sim_alg} in parallel through \emph{spatial loop unrolling}. The data access and communication patterns across the multiple PEs are determined by the \emph{dataflow} scheme~\cite{eyeriss_isca16}. Typically, the dataflow is carefully orchestrated so that data accesses to more expensive memories, including the storage in other PEs and the large shared buffers, can be minimized. We will provide a comprehensive dataflow taxonomy in Section~\ref{sec:dataflow}. For now we use stationary characteristics from \cite{eyeriss_isca16} as the dataflow labels, such as weight stationary (WS), row stationary (RS), etc., to represent the choices in the dataflow dimension in Figure~\ref{fig:3d_space}. Here we assume dataflow choice is purely a subset of mapping space, which is orthogonal to the underlying architecture. Later we will provide more explanations about this assumption.

\textbf{Resource Allocation:} Hardware resource allocations, such as the dimensions of the PE array and the size of each level in the memory hierarchy, are also essential to the performance and efficiency of the accelerator. They determine the computation throughput, the location of the data, and the energy cost and latency for each memory access. For example, since the energy cost and latency of each data access grow with memory size, an efficient design needs to carefully size each memory level to optimally balance buffering sufficient data for high locality while being as small as possible to minimize fetch energy. In summary, the resource allocation axis needs to cover various hardware choices. Here in Figure~\ref{fig:3d_space} we mainly consider the number of MAC units $N$, and the memory size $S_i$ at each level $i$, represented as a  vector $(N, S_1, S_2, \ldots)$.

\textbf{Loop Blocking:} 
Assuming a multi-level memory hierarchy (e.g., register files, on-chip SRAM, and off-chip DRAM), we want to schedule the computation to maximize the data reuse in the near, smaller memories to lower overall energy cost. 
Since all the data fetched\,---\,inputs, outputs, and weights\,---\,can be potentially reused,
an optimal schedule must choose the best data reuse opportunities. The techniques of loop blocking and reordering, which together we refer to as loop blocking~\cite{blocking_cnn_arxiv16}, transforms the seven nested loops in Algorithm~\ref{alg:sim_alg} into a much larger number of loops, and generates different data access/reuse  patterns for each memory level. 

Using these three factors enables us to create a design space that is comprehensive and systematic, enabling us to project existing DNN accelerators as shown in Figure~\ref{fig:3d_space}. We do not explicitly show the loop blocking schemes for each architecture in this figure, since most prior work did not report their loop blocking strategy, or simply exhaustively searched for an ad-hoc optimized scheme. This figure clearly shows the wide design space current accelerators occupy. 



\subsection{A Formal Dataflow Taxonomy}
\label{sec:dataflow}

Since blocking is by definition operations on loop nests, we next show how the dataflow of an accelerator can also be represented by loop operations. Noticing the connection between dataflow and spatial loop unrolling, we represent the dataflow of an accelerator through the mapping of particular loops to the parallel computation structures, similar to the terminology for the spatial partitioning in Timeloop~\cite{timeloop}. In other words, the data communication pattern is determined by \emph{which loops are spatially unrolled} in hardware, and which are not. For example, if the $X$ and $Y$ loops are unrolled onto the 2D array, then each PE produces a single output pixel. This output stationary pattern implies that input pixels will be reused across neighbor PEs as they contribute to multiple output pixels in a convolution, and the filter weights shared by all output pixels must be transferred to all PEs. If we instead unroll the $F_X$ and $F_Y$ loops, we obtain a weight stationary pattern, where the weights stay and are reused within the same PEs, but the inputs and outputs are spatially broadcast or accumulated.

To concisely represent dataflows using spatial loop unrolling schemes on 2D PE arrays, we use the syntax $U\!\mid\!V$, where $U$ and $V$ denote the loops unrolled across the vertical and horizontal dimensions, respectively. 
Given $L$-level (excluding those with loop bounds as 1) nested loops in the algorithm and $d$ spatial dimensions in the accelerator, there are $\binom{L}{d}$ possible dataflow choices. 
For a 2D array and the unblocked algorithm, the number of dataflow types is $\binom{7}{2} = 21$ for a CONV layer, and $\binom{3}{2} = 3$ for a fully-connected layer.


\begin{figure}
    \vspace{-10pt}
    \centering
    \subfloat[No replication.]{
        \centering
        \includegraphics[width=0.22\textwidth]{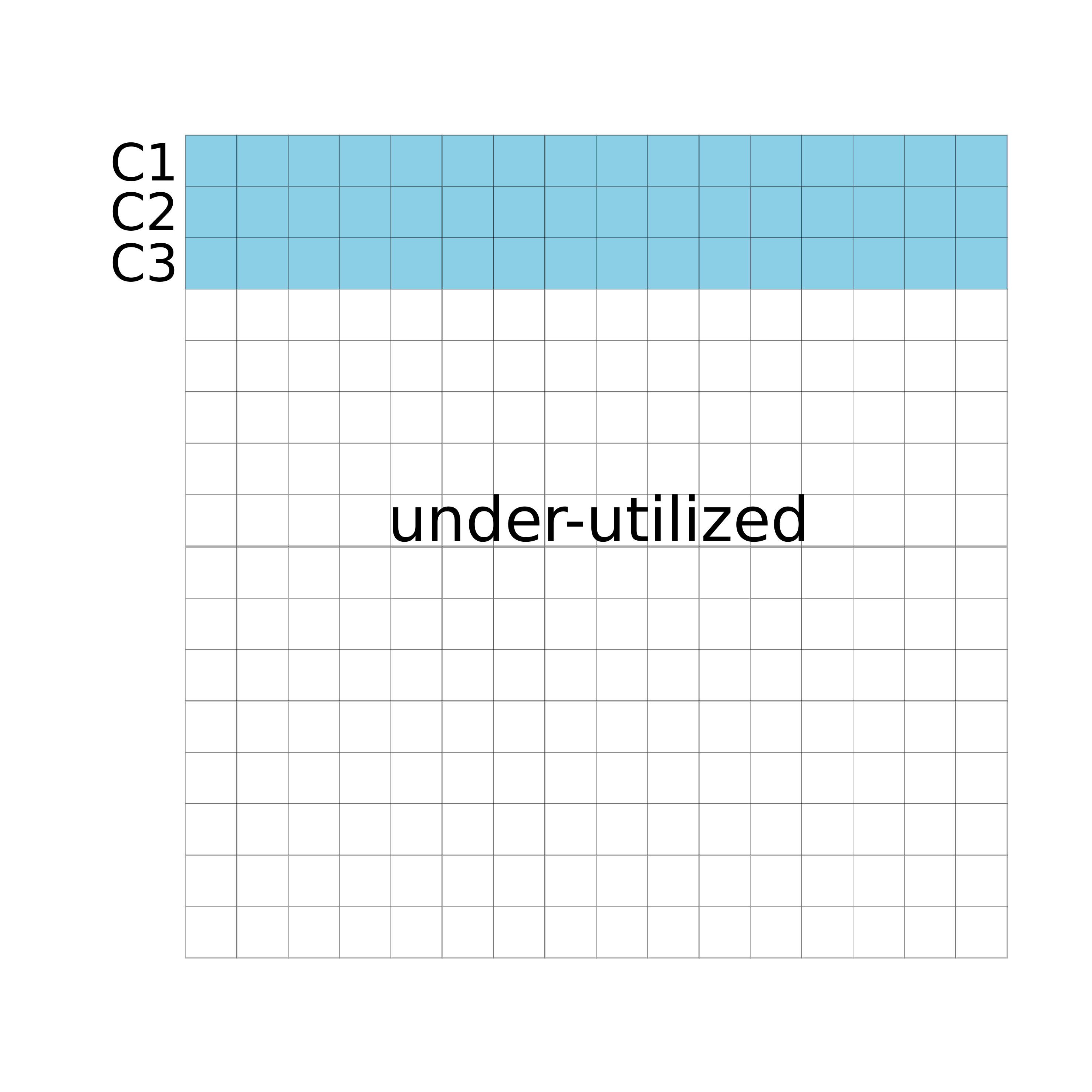}
        \label{fig:no-replication}
    }
    \subfloat[With replication.]{
        \centering
        \includegraphics[width=0.22\textwidth]{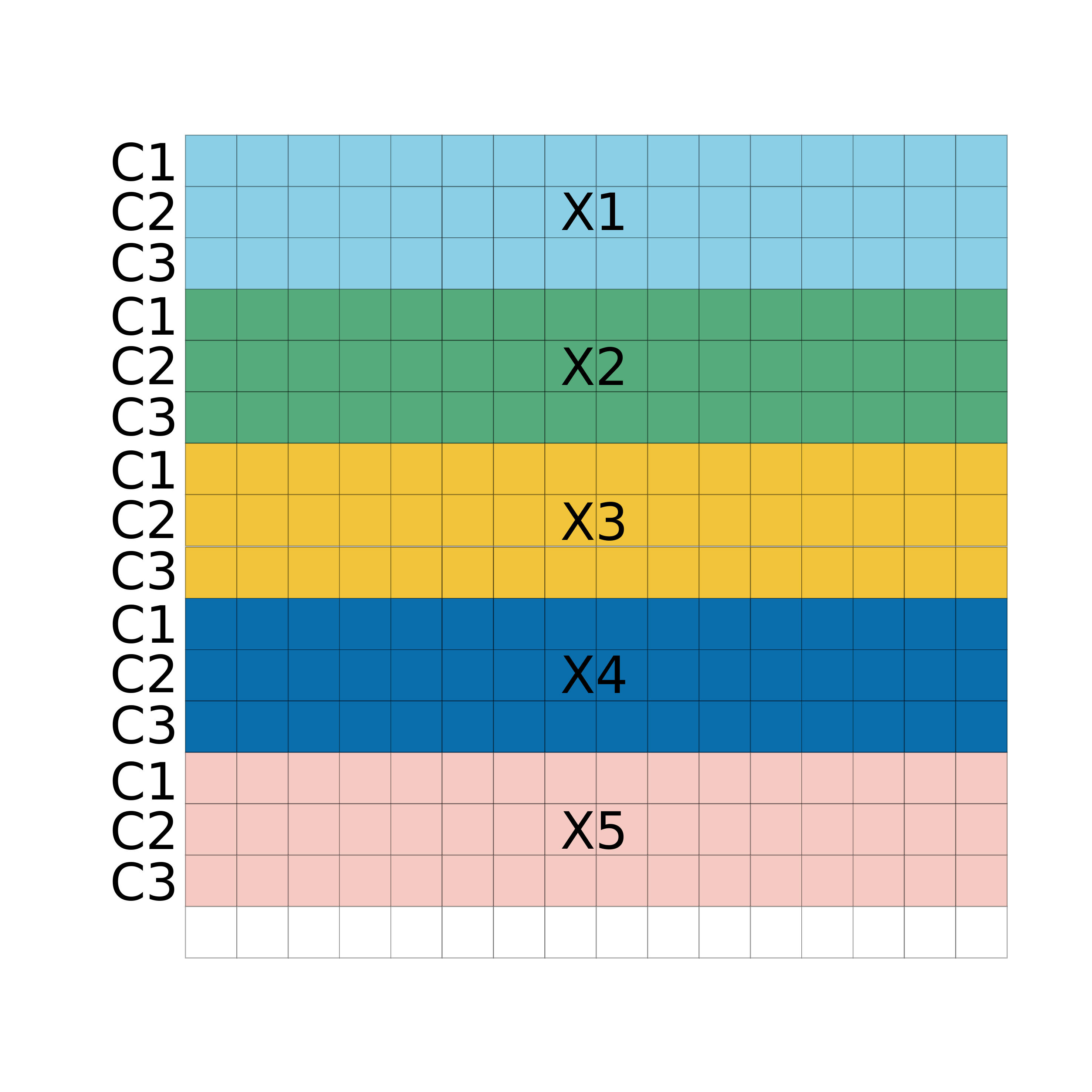}
        \label{fig:with-replication}
    }
    \caption{Computation resource utilization can be improved by replication.}
    \label{fig:replic}
\end{figure}

However, the above considered dataflows, which unroll a single loop at each spatial dimension, can potentially result in under-utilizating computation resources. For instance, as illustrated in Figure~\ref{fig:no-replication}, when unrolling a loop $C$ with size of 3 on the vertical dimension of a 16$\times$16 PE array, only 3 of the 16 rows of PEs are utilized, leaving the remaining PEs idle. To overcome this issue, in addition to unrolling loop $C$, another loop such as $X$ is also unrolled by a factor of 5, as shown in Figure~\ref{fig:with-replication}, improving the utilization ratio from $3/16$ to $15/16$. Therefore, it is of great importance to support unrolling multiple loops onto one spatial dimension. This improvement is called \emph{replication}~\cite{eyeriss_isca16, chen2018eyeriss}, which processes multiple small loops in parallel to increase resource utilization. Our loop-based taxonomy can also nicely and consistently express it using $U\!\mid\!VW$ or $UW\!\mid\!V$, depending on the replicated dimension. When replication is supported, the number of dataflow choices for a layer further increases. 

\begin{table}
\centering
\caption{Common dataflows from \cite{eyeriss_isca16} expressed using spatially unrolled loops.}  
\label{tab:table_dataflows}
\begin{threeparttable}
\begin{tabular}{cc}
    \toprule
    \textbf{Dataflow} & \textbf{Representation} \\
    \midrule
    Output stationary & $X\!\mid\!Y$ \\
    Weight stationary & $F_X\!\mid\!F_Y$ \\
    Row stationary    & $F_Y\!\mid\!Y$ \\
    Weight stationary & $C\!\mid\!K$ \\
    \bottomrule
\end{tabular}
\end{threeparttable}
\end{table}

\begin{figure}
    \centering
    \includegraphics[width=0.45\textwidth]{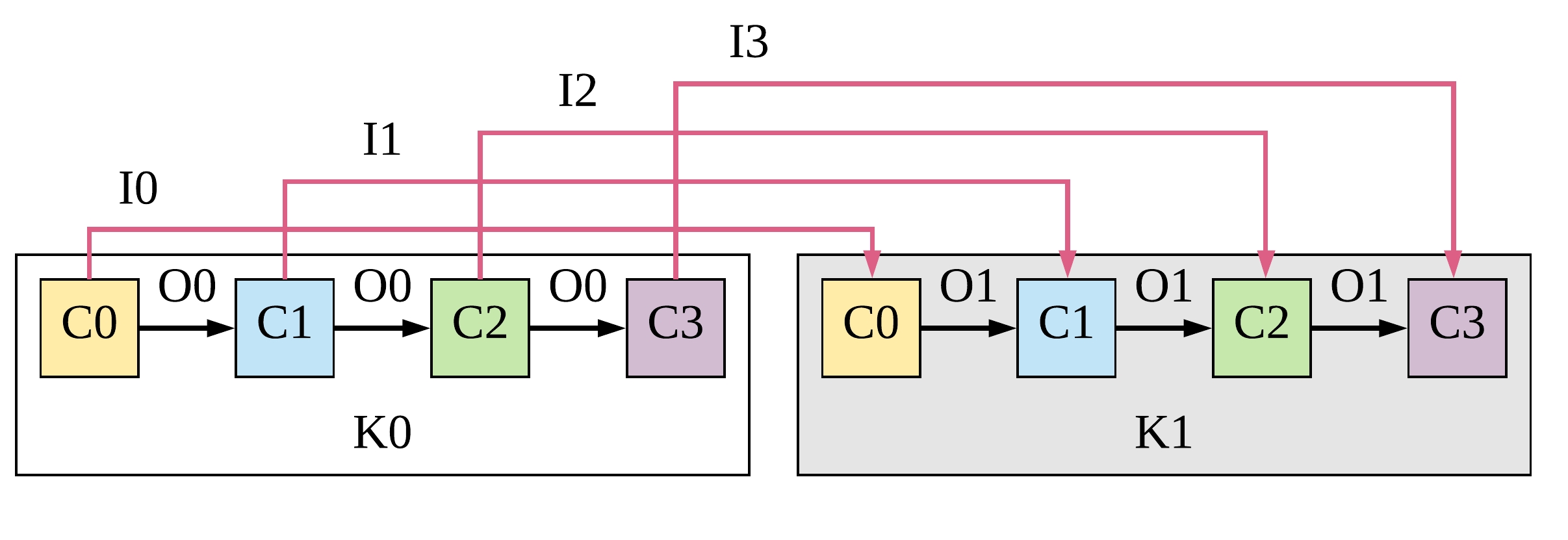} 
    \caption{Loops $C$ and $K$ are unrolled onto a 1D array with dataflow $CK$. Outputs are communicated between adjacent PEs, while inputs are communicated across groups.}
    \label{fig:replication}
\end{figure}

Note that with replication (i.e. mapping multiple loops onto the same physical dimension), the data communication pattern is no longer uniform: intra-loop data can be communicated among nearest-neighbor PEs, while inter-loop data have to be sent multiple hops away with higher communication cost. Syntactically, we represent this by ordering the loops mapped to the same dimension, where the PEs generated by unrolled loops to the left have shorter communication distances than the loops on the right. Figure~\ref{fig:replication} shows an example of unrolling both $C$ and $K$ loops onto a 1D array. The eight PEs have been divided into two groups, each working on a output channel $K_i$. Within each group, different PEs process different input channels $C_i$. The outputs are only communicated among the nearest PEs, while the inputs have to transfer from one group to the other with a cost four times that of nearest neighbor communication. As was mentioned, Eyeriss V2~\cite{chen2018eyeriss}, enables flexible replication by providing  flexible interconnections between the processing elements. Fortunately, the interconnects provided by some hardware targets, including FPGAs and CGRAs, naturally provide this flexibility.

\textbf{Advantages:}
The loop-based approach builds upon the ideas of stationary characteristics~\cite{eyeriss_isca16}, and provides a more precise definition of each dataflow while expanding the range of flows that can be described.  Table~\ref{tab:table_dataflows} shows several common dataflows in prior work expressed using our taxonomy. As an example, $C\!\mid\!K$ is a widely adopted dataflow (Figure~\ref{fig:3d_space}, and \cite{fused_layer_cnn_micro16, cnn_fpga_util_fpl16, cnn_fpga_util_isca17, suda2016throughput}) due to its flexibility to also map matrix multiplications in MLPs and LSTMs. Even though $C\!\mid\!K$ also keeps the weight stationary in PEs, its data reuse pattern is quite different from the weight-stationary dataflow $F_X\!\mid\!F_Y$ introduced by \cite{eyeriss_isca16}. Furthermore, more complicated dataflows, such as a hybrid weight and output stationary pattern, are easy to represent in our taxonomy, e.g., $C\!\mid\!KX$, demonstrating its completeness. 


Using the loop-based dataflow taxonomy, dataflows and loop blocking schemes can now both be expressed as transformations of the seven nested loops in Algorithm~\ref{alg:sim_alg}. There are many existing approaches to find loop transformations that optimize some cost functions, in order to either realize optimal software implementations on CPUs or GPUs, or generate optimal hardware designs using FPGAs or ASICs. These include general approaches like Polyhedral analysis~\cite{Mullapudi:2015:PAO:2694344.2694364, Zuo:2013:IHL:2435264.2435271} and studies specific to DNNs~\cite{blocking_cnn_arxiv16, kwon2019understanding}. 

In the next section, we show that these loop transformations can be expressed in Halide's scheduling language, and we use this language to specify the schedules and hardware resources (both memory and compute units) of any specific DNN accelerator design. Although Halide and polyhedral models both support affine loop transformations, we choose Halide over creating a new system with a polyhedral model, in order to leverage Halide’s compact scheduling primitives and mature compilation framework. This allows us to separate the hard optimization problem -- finding the right schedule, from the mechanical transformation and manual effort to implement it. 


%% file: implementation.tex
\vspace{-1em}
\section{Halide Accelerator Design}
\label{sec:halideHardware}

Halide~\cite{halide2013} is a domain-specific language (DSL), originally designed for image processing but generally applicable to dense loop-structured computation including linear algebra and DNNs. The key idea in Halide is to split the computation to be performed (the \emph{algorithm}) from the order in which it is done (the \emph{schedule}). To allow the user to express the implementation order, Halide provides a compact and elegant language to represent loop transformations.  These transformations along with commands that create intermediate storage are sufficient to completely specify a DNN accelerator, and the blocking of the algorithm which most efficiently utilizes it. 


\textbf{Halide algorithm:}
Halide represents the computation algorithm in pure functional form. The following example shows the Halide algorithm for a CONV layer with $3\!\times\!5\!\times\!5$ filter size.
\begin{lstlisting}[language=C++, morekeywords={RDom}]
// To perform a 5 x 5 convolution with 3 channels
// RDom(xMin, xExt, yMin, yExt, kMin, kExt)
RDom r(-2, 5, -2, 5, 0, 3);
output(x, y, k) += input(x + r.x, y + r.y, r.z) 
                  * w(r.x + 2, r.y + 2, r.z, k);
\end{lstlisting}
The \texttt{RDom} keyword defines a multi-dimensional reduction domain, over which an iterative computation such as a summation is performed. A \texttt{RDom} is defined by the minimum position and extent in each of its dimensions. In the CONV layer example, the \texttt{RDom} covers the width and height of the filters and the number of input fmaps, over which the accumulation will iterate.

While the algorithm defines the functionality of the computation, it does not specify the ordering of parallel operations or data accesses. These are further controlled using Halide schedules, which consist of \emph{scheduling primitives} applied to various stages of the algorithm. These primitives can express the required loop blocking, resource allocation, and, with minor extensions, dataflow choices as well. 

\subsection{Halide Schedules}

With only small extensions to the current Halide scheduling language, we can explore all hardware architectures and software scheduling choices in the design space introduced in Section~\ref{sec:design_space}. Table~\ref{tab:table_dimension} summarizes how the scheduling primitives control each of the three design space dimensions. Listing~\ref{list:example-schedule} shows an example Halide schedule for the above Halide algorithm of a CONV layer. 

\begin{table}
\centering
\caption{Halide scheduling primitives that control each dimension of the 3D design space.}  
\label{tab:table_dimension}
\begin{threeparttable}
\begin{tabular}{lc}
    \toprule
    \textbf{Dimensions} & \textbf{Scheduling primitives} \\
    \midrule
    Loop blocking & \texttt{split}, \texttt{reorder}   \\
    Resource allocation & \texttt{in}, \texttt{compute\_at} \\
    Dataflow & \texttt{unroll}, \texttt{systolic} \\
    \midrule
    Overall scope & \texttt{accelerate} \\
    \bottomrule
\end{tabular}
\end{threeparttable}
\end{table}

\begin{lstlisting}[language=C++,morekeywords={in, split, reorder, accelerate, compute_at, update, systolic, unroll, Var},
float,
label=list:example-schedule,
caption={An example Halide schedule for the CONV layer algorithm.}]
Var xo, yo, xi, yi;
output.update().split(x, xo, xi, 8)
               .split(y, yo, yi, 8) 
               .reorder(xi, yi, xo, yo); 
input.in().compute_at(output, xo);
w.in().compute_at(output, xo);
output.accelerate({input, w});
output.update().unroll(xi, 4);
output.update().systolic({xi});
\end{lstlisting}

\begin{figure}
    \centering
    \includegraphics[width=0.4\textwidth]{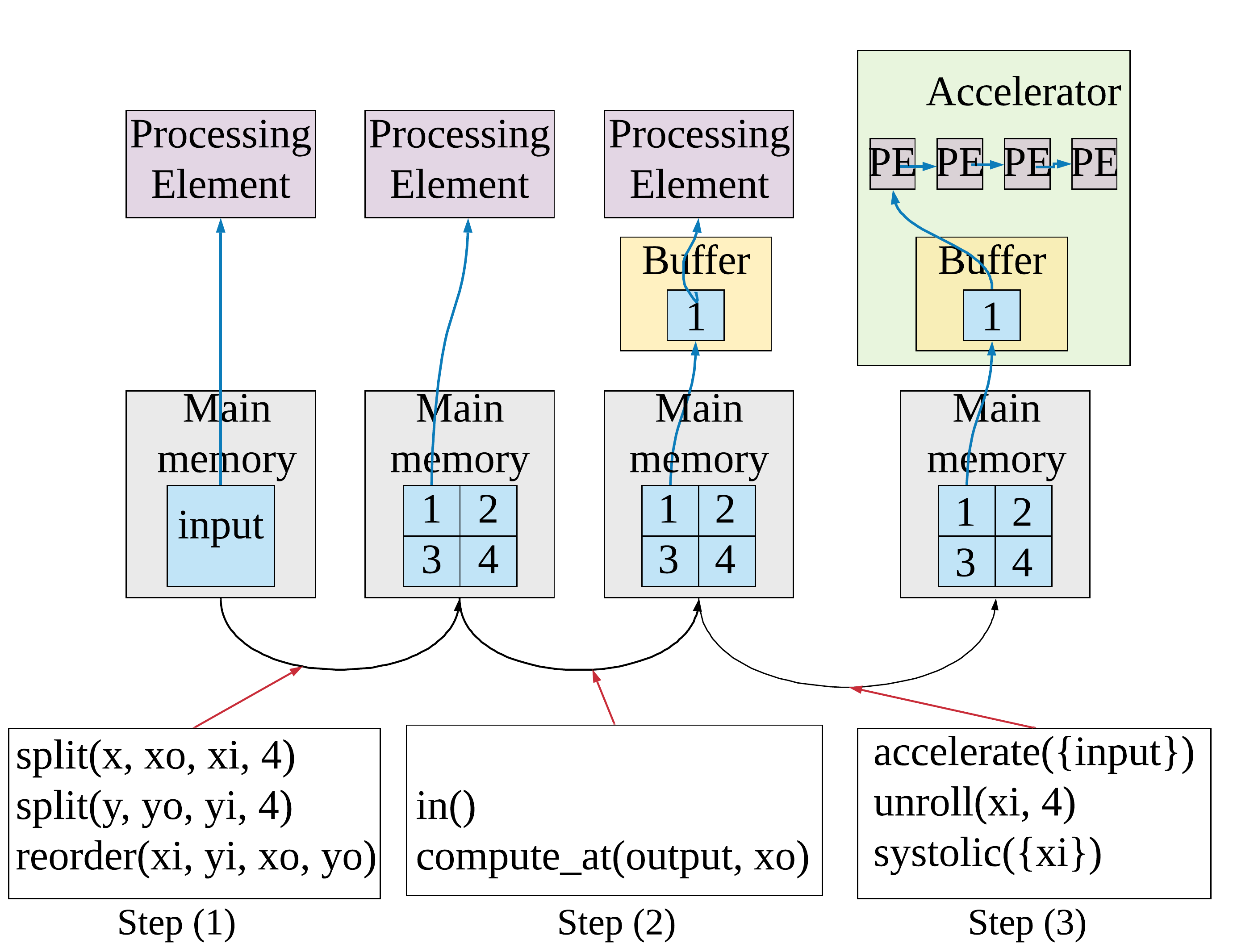}
    \caption{The initial design fetched the data as one large block from memory.  After the split and reorder, the data is broken into 4 smaller tiles. Next a local buffer for one tile is allocated, and finally a 4 PE systolic array is implemented to process the data.
    }
    \label{fig:halide_schedule}
\end{figure}

The algorithm and schedule provide a user-facing language to construct hardware. Figure~\ref{fig:halide_schedule} pictorially shows this example schedule in details. From left to right, we iteratively apply three sets of scheduling primitives to achieve the final accelerator structure. Listing~\ref{list:generated-ir} presents the intermediate representation (IR) generated by Halide, corresponding to the third phase in Figure~\ref{fig:halide_schedule}. The rest of this section describes the three transformations in that figure.

\begin{lstlisting}[language=C++,morekeywords={alloc}, 
float,
label=list:generated-ir,
caption={Intermediate representation generated by Halide after using \texttt{in} and \texttt{compute\_at} combined with \texttt{split} and \texttt{reorder}, corresponding to the third phase after step (2) in Figure~\ref{fig:halide_schedule}.}]
// To generate output of size 16 x 16 x 64 
for (k, 0, 64)
  for (yo, 0, 2)
    for (xo, 0, 2)
      // Allocate local buffer for input.
      alloc ibuf[8 + 5 - 2, 8 + 5 - 2, 3]
      // Copy input to buffer.
      ibuf[...] = input[...]
      // Allocate local buffer for w.
      alloc wbuf[5, 5, 3, 1]
      // Copy w to buffer.
      wbuf[...] = w[...]
      for (yi, 0, 8)
        for (xi, 0, 8)
          for (r.z, 0, 3)
            for (r.y, -2, 5)
              for (r.x, -2, 5)   
                output(xi, yi, 0) +=
                  ibuf(xi + r.x, yi + r.y, r.z)
                  * wbuf(r.x + 2, r.y + 2, r.z, 0)
                        
\end{lstlisting}

\textbf{Loop blocking:}
The existing Halide scheduling primitives are primarily designed for loop transformations on general-purpose processors, but the syntax and semantics also support loop blocking on accelerators thanks to the same underlying principles. Lines 2--4 of Listing~\ref{list:example-schedule} use \texttt{split} to break the \texttt{x} and \texttt{y} loops into two levels, where the inner loops have 8 iterations. \texttt{split} can also be applied repeatedly to create more levels. \texttt{reorder} interchanges the loops, setting the order of computation and data access. These two primitives can realize different \emph{loop blocking} schemes, splitting the data into multiple smaller subtiles (4 tiles of 8$\times$8 in this example) that are processed in a certain order (\texttt{x} then \texttt{y}). Step (1) in Figure~\ref{fig:halide_schedule} visually shows the four tiles that are created due to the new loops (lines 3--4 and 13--14 in Listing~\ref{list:generated-ir}).

\begin{figure}
    \centering
    \subfloat[Systolic array.]{
        \centering
        \includegraphics[width=0.22\textwidth]{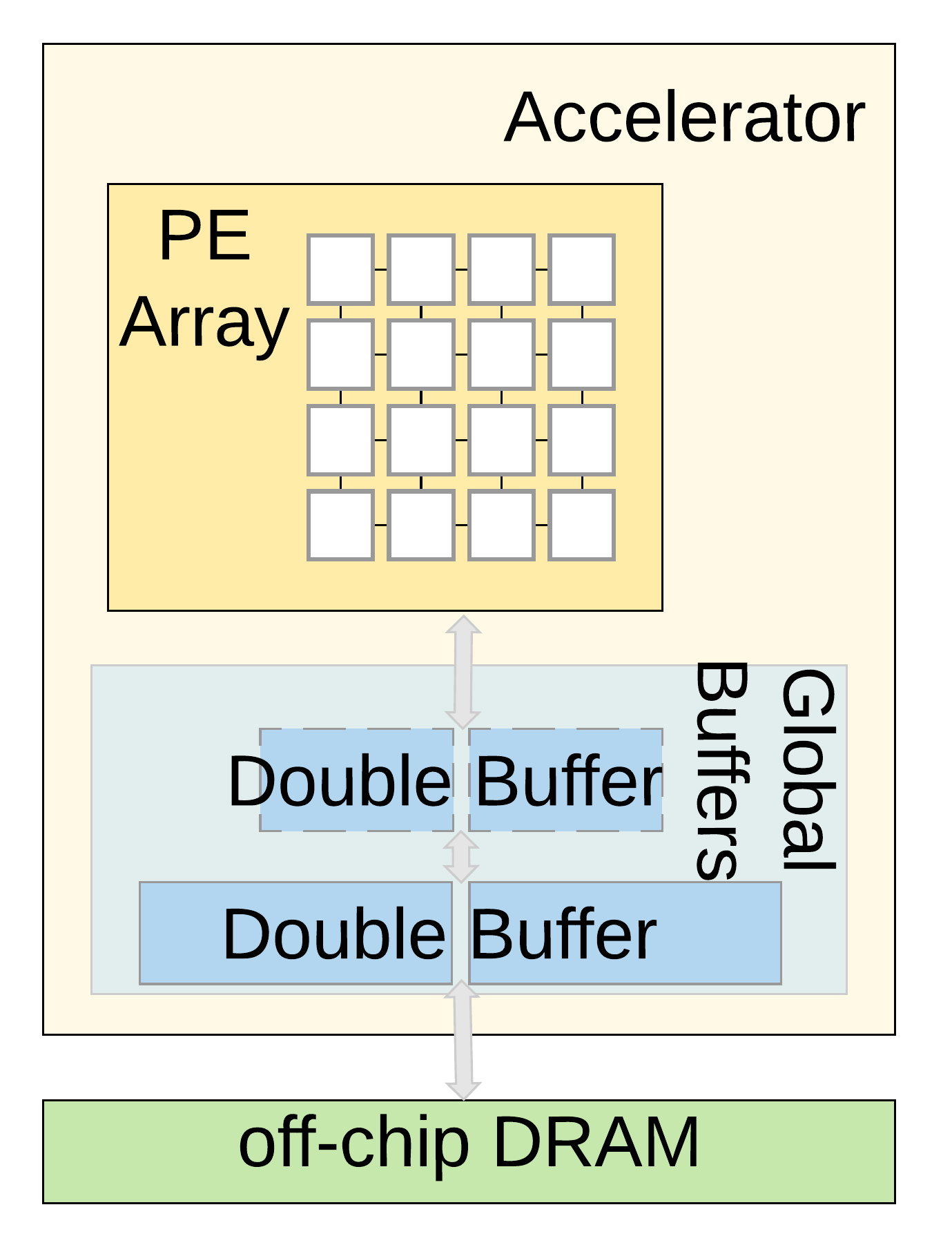}
        \label{fig:db_template:systolic}
    }
    \subfloat[Reduction tree.]{
        \centering
        \includegraphics[width=0.22\textwidth]{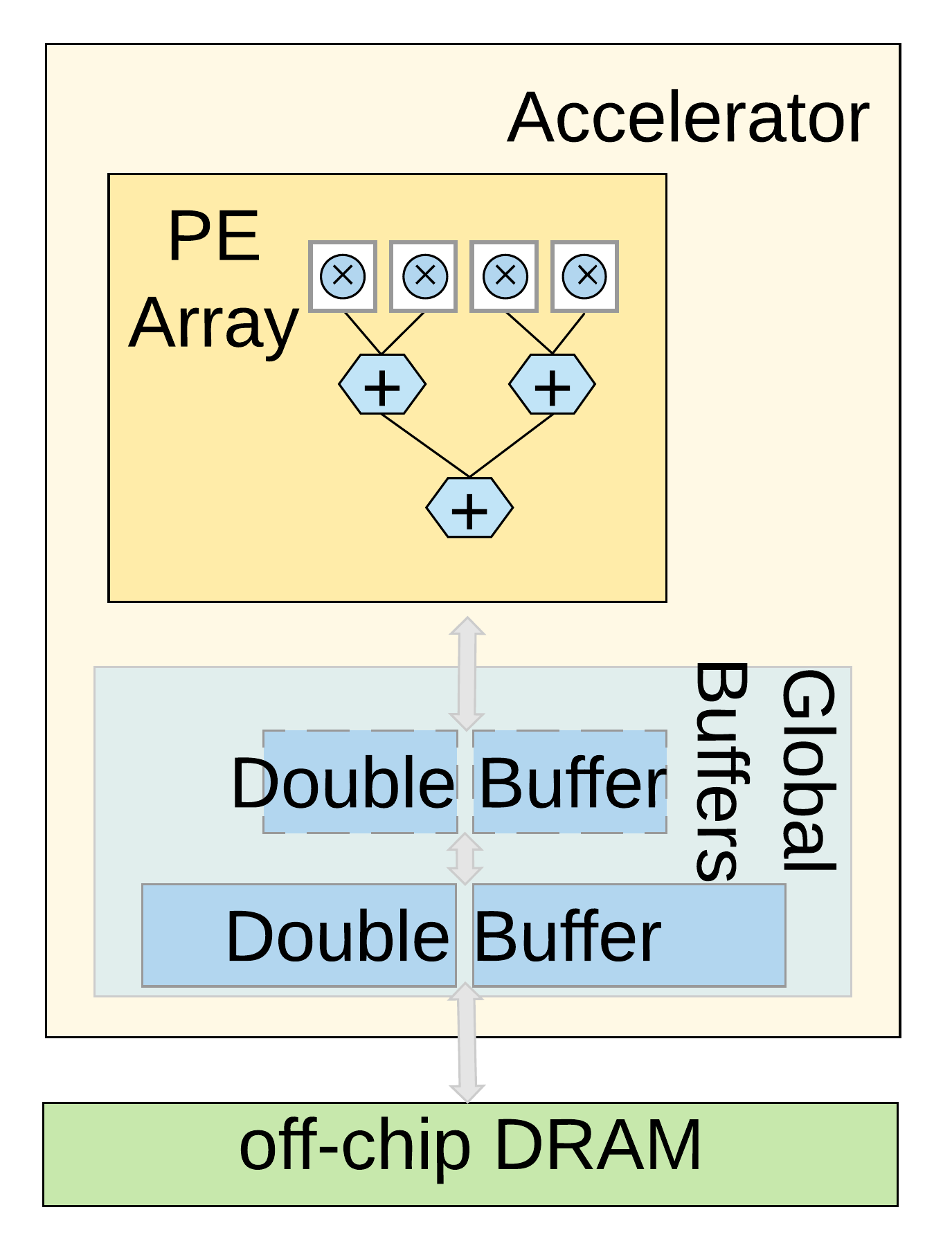}
        \label{fig:db_template:reduction_tree}
    }
    \caption{DNN accelerator architecture consisting of a PE array and a memory hierarchy.}
    \label{fig:db_template}
\end{figure}

\textbf{Resource allocation of memory hierarchy:}
After splitting the data, we need to allocate SRAM buffer resources so that each data subtile can be cached on-chip while being processed to reduce their access cost. The existing primitives \texttt{in} and \texttt{compute\_at} (lines 5--6 of Listing~\ref{list:example-schedule}) together introduce additional memory levels, and specify at which loop iteration to fetch which subtiles into the buffers (Figure~\ref{fig:halide_schedule}). The compiler combines this information with loop sizes, and instantiates the correct number of memory levels with appropriate size and data layout for each buffer. As shown by lines 5--12 in  Listing~\ref{list:generated-ir}, by calling \texttt{in} and \texttt{compute\_at} together for both \texttt{input} and \texttt{w}, two local buffers with required sizes are allocated within loop \texttt{xo} for input and weight data respectively. This allows us to explore different \emph{resource allocation} choices. For each memory level, we use a \emph{double buffer} design (Figure~\ref{fig:db_template}), which enables overlapping computation and data fetch: during the computation of the current tile, the next tile is loaded into the alternate buffer. 

\begin{figure*}
    \centering
    \subfloat[A systolic dataflow $F_Y\!\mid\!Y$, generated by unrolling the output fmap height $Y$ and filter weight height $F_Y$.]{
        \centering
        \includegraphics[width=0.25\textwidth]{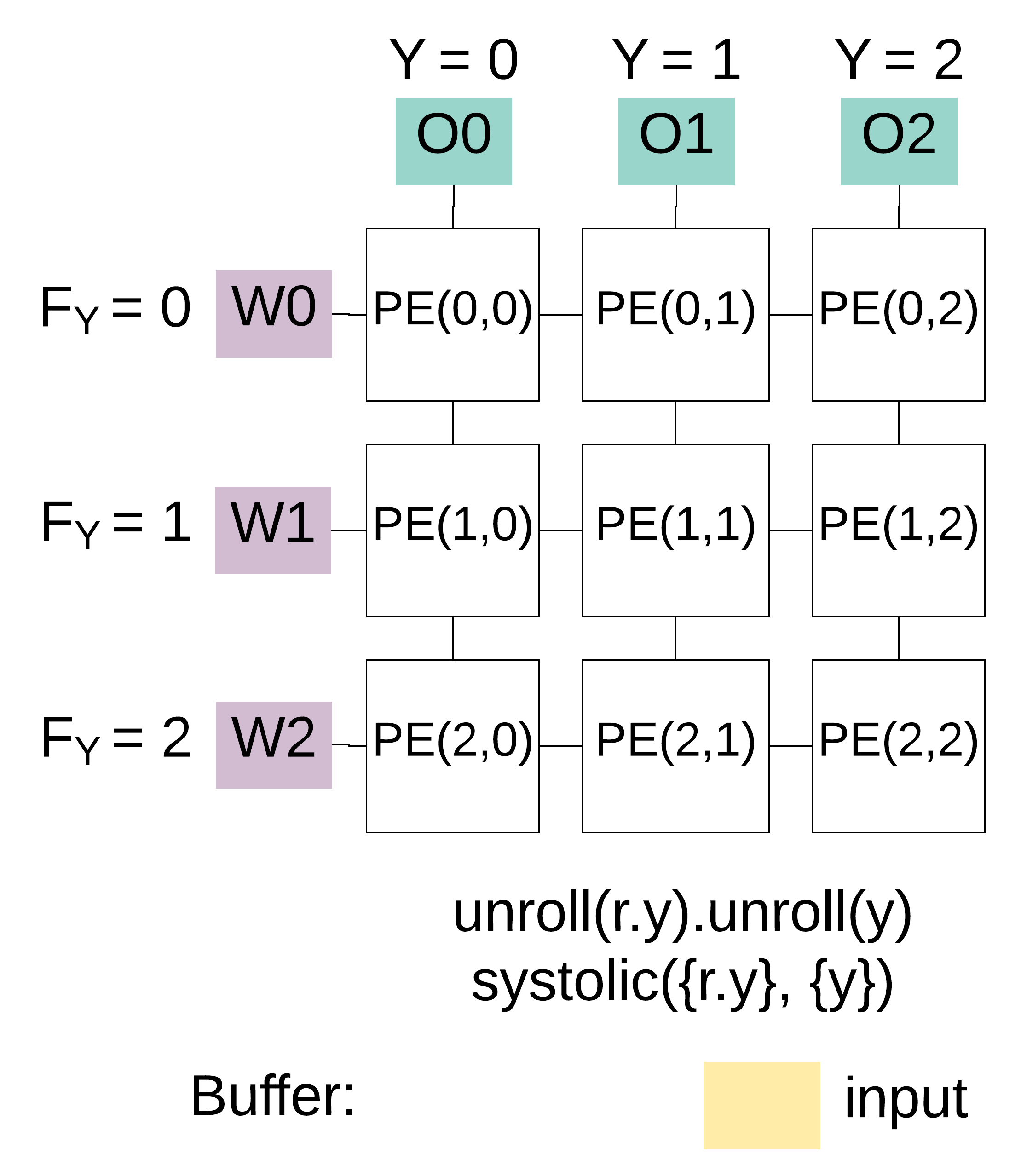} 
        \label{fig:eyeriss_dataflow}
    }
    \hspace{1em}
    \subfloat[A systolic dataflow $C\!\mid\!K$, generated by unrolling the input and output fmap dimensions $C$ and $K$.]{
        \centering
        \includegraphics[width=0.25\textwidth]{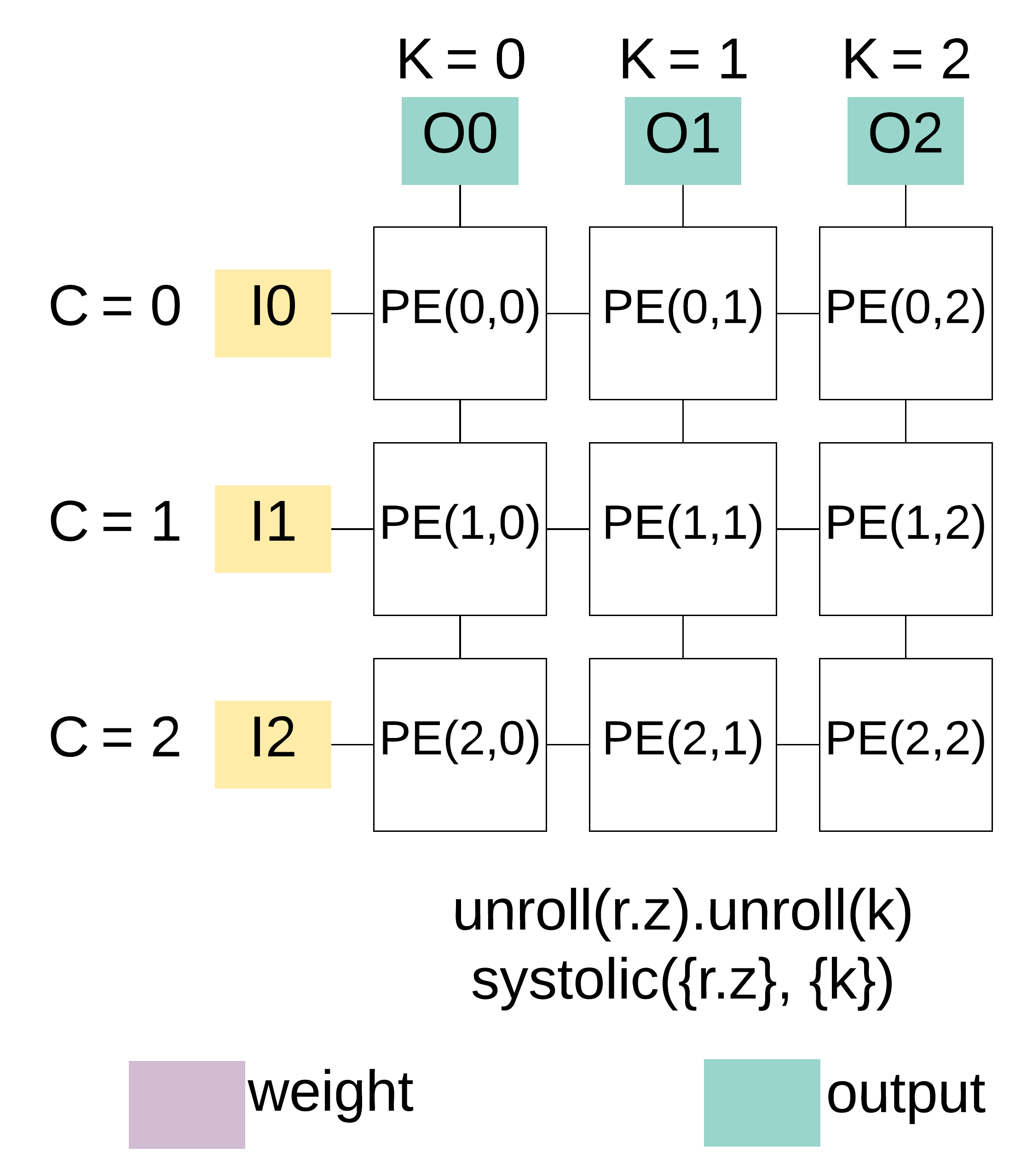} 
        \label{fig:tpu_dataflow}
    }
    \hspace{1em}
    \subfloat[A reduction tree of dataflow $C$, generated by unrolling the input fmap dimension $C$.]{
        \centering
        \includegraphics[width=0.25\textwidth]{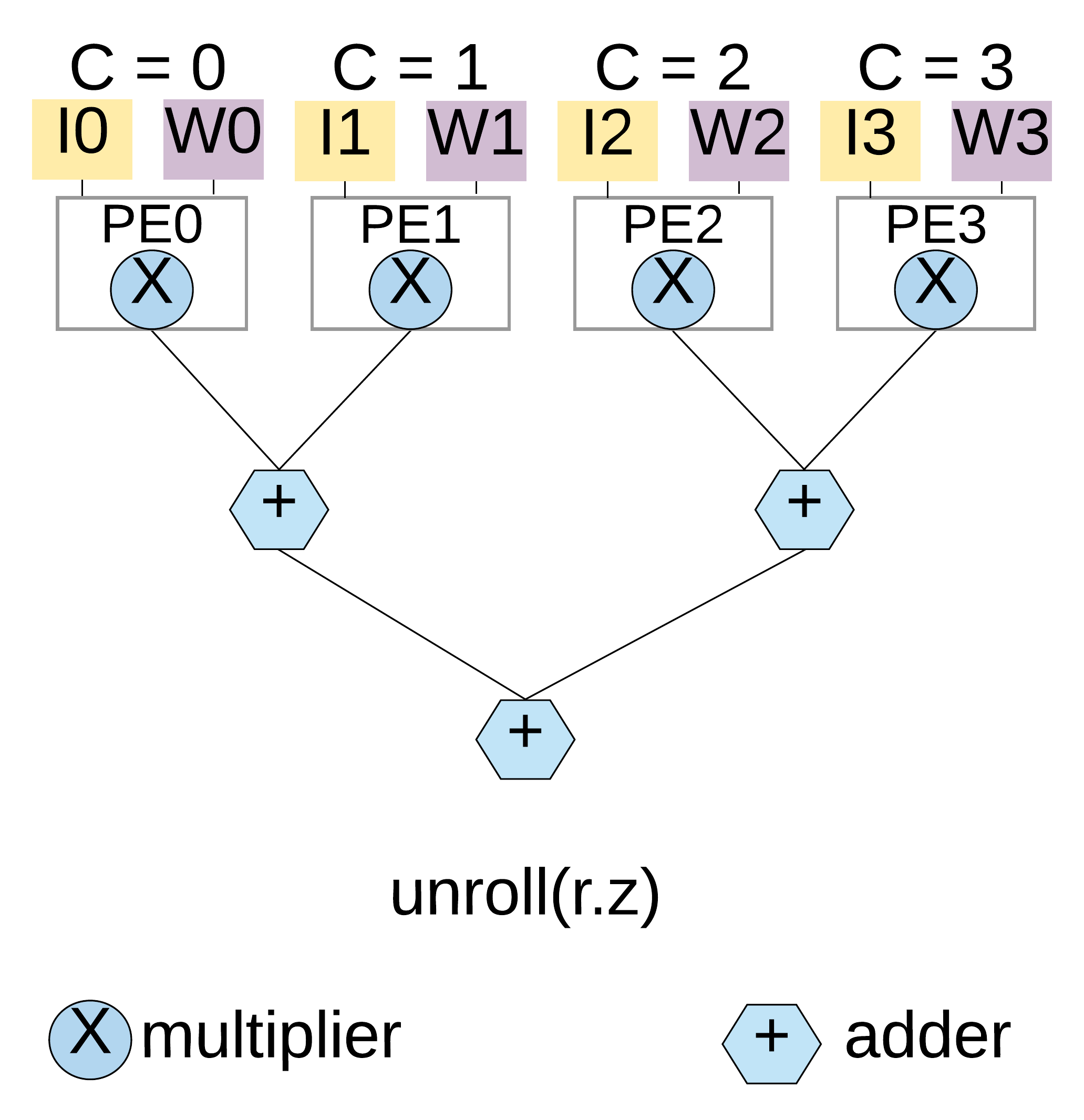} 
        \label{fig:reduction_tree}
    }
    \caption{Different PE array micro-architectures generated from Halide scheduling primitives.}
    \label{fig:dataflow_impl}
\end{figure*}

\textbf{Dataflow and PE array micro-architecture:}
While the previous two dimensions can be covered using existing Halide primitives, expressing \emph{dataflow} requires extensions to support the complex on-chip data propagation patterns uniquely appearing in accelerators. First, like Pu et al.~\cite{halidefpga2017}, we overload the existing \texttt{unroll} primitive to specify spatial loop unrolling onto the PE array. As discussed in Section~\ref{sec:dataflow}, given a 1D dataflow $U$ or 2D dataflow $U\!\mid\!V$, we spatially unroll the loops $U$ and $V$ on each physical dimension, respectively. For example, in Figure~\ref{fig:halide_schedule} step (3), the loop \texttt{xi} is unrolled to execute in parallel on 4 systolic PEs.

Second, we support various types of PE array micro-archi\-tectures. We introduce a new primitive, \texttt{systolic}, which realizes a systolic PE array as shown in Figure~\ref{fig:db_template:systolic}, and allows direct inter-PE data communication without always fetching data from higher-level data buffers~\cite{eyeriss_isca16, tpu_isca17}. Combined with different \texttt{unroll} primitives, we can realize different dataflows on the PE array. Figure~\ref{fig:eyeriss_dataflow} maps the $F_Y\!\mid\!Y$ dataflow used in Eyeriss~\cite{eyeriss_isca16}, by unrolling the $F_Y$ and $Y$ loops. It transfers multiple rows of filter weights horizontally, and accumulates multiple columns of output fmaps vertically. Alternatively, Figure~\ref{fig:tpu_dataflow} performs matrix multiplications using dataflow $C\!\mid\!K$, which is used by a large group of designs including Google's TPU~\cite{tpu_isca17}.

Without applying \texttt{systolic}, the PEs are by default organized into reduction tree structures~\cite{chen2014diannao}, as shown in Figure~\ref{fig:db_template:reduction_tree}. Figure~\ref{fig:reduction_tree} provides one example dataflow on a 1D reduction tree, which unrolls the loop $C$ to multiply input pixels from different input fmaps with their corresponding weights, and accumulates the products into a single output pixel in an output fmap.

Using the two micro-architectures in Figure~\ref{fig:db_template} as building blocks, we can generate a variety of accelerator designs by composition. They can be described by applying \texttt{unroll} at different loop levels, and calling \texttt{systolic} at the corresponding unrolled loops. For instance, we can have a reduction tree of PEs acting as a node in a systolic array, or multi-level reduction trees, effectively supporting a wide range of designs including ARM ML processor~\cite{ARM} and NVDLA~\cite{Nivida}. We could additionally introduce new primitives similar to \texttt{systolic}, to support other PE array micro-architectures if desired.

\textbf{Accelerator scope:}
Finally, we introduce an additional primitive, \texttt{accelerate}, which defines the scope of the hardware accelerator and the interface to the rest of the system, in a similar manner to Pu et al.~\cite{halidefpga2017}.

\subsection{Implementation}

Given this concise and concrete expression for DNN accelerators using Halide schedules, we extended the Halide compiler to generate hardware from these descriptions. 
As a result, different DNN accelerator designs and mapping schemes can be realized by simply changing the Halide schedule associated with the same Halide algorithm. We can also use it to easily recreate previously proposed designs for a fair comparison (see Section~\ref{sec:result}).

Our implementation of the Halide toolchain is built on top of Pu's work~\cite{halidefpga2017}, which was designed for generating hardware for image processing pipelines. To create our system, we needed to extend this work in three important ways.  The first was to add support for systolic arrays, which was challenging due to the diversity of the PE connectivity and the complexity of creating a state machine for the control logic inside each PE. Next, hardware virtualization was added to map different layers in a DNN (represented as different Halide functions) onto the same hardware module. Otherwise, na\"ively instantiating spatially separated modules for each layer in large DNNs would consume unrealistically large silicon area. Finally, we extended Pu's work to generate ASIC as well as FPGA implementations.

%% file: method.tex
\section{Methodology}
\label{sec:method}

Our Halide-based accelerator design flow in Section~\ref{sec:halideHardware} supports both FPGA and ASIC backends. The FPGA results enabled us to validate that the system was functional and produced efficient designs: it achieved similar GOPs and DSP utilization when compared against manually optimized designs~\cite{qiu2016going, suda2016throughput}.

\textbf{ASIC Hardware synthesis toolchain:} 
Our extended Ha\-lide compiler generates C++ code specialized for Catapult High-Level Synthesis, which is then compiled to RTL designs in Verilog. We synthesize the RTL designs in a 28\,nm technology using Synopsys Design Compiler. Standard cells and memory models from commercial vendors are used for power, performance, and area analysis. We use 16-bit arithmetic for inference tasks throughout this paper. All of our ASIC designs achieve 400\,MHz frequency with no timing violations. For power analysis, the appropriate switching activities are set on all the primary ports and propagated through the design using the design tools.


\begin{figure*}
    \centering
    \subfloat[Energy breakdown comparison between actual synthesized designs and the analytical model.]{
        \centering
        \includegraphics[width=0.45\textwidth]{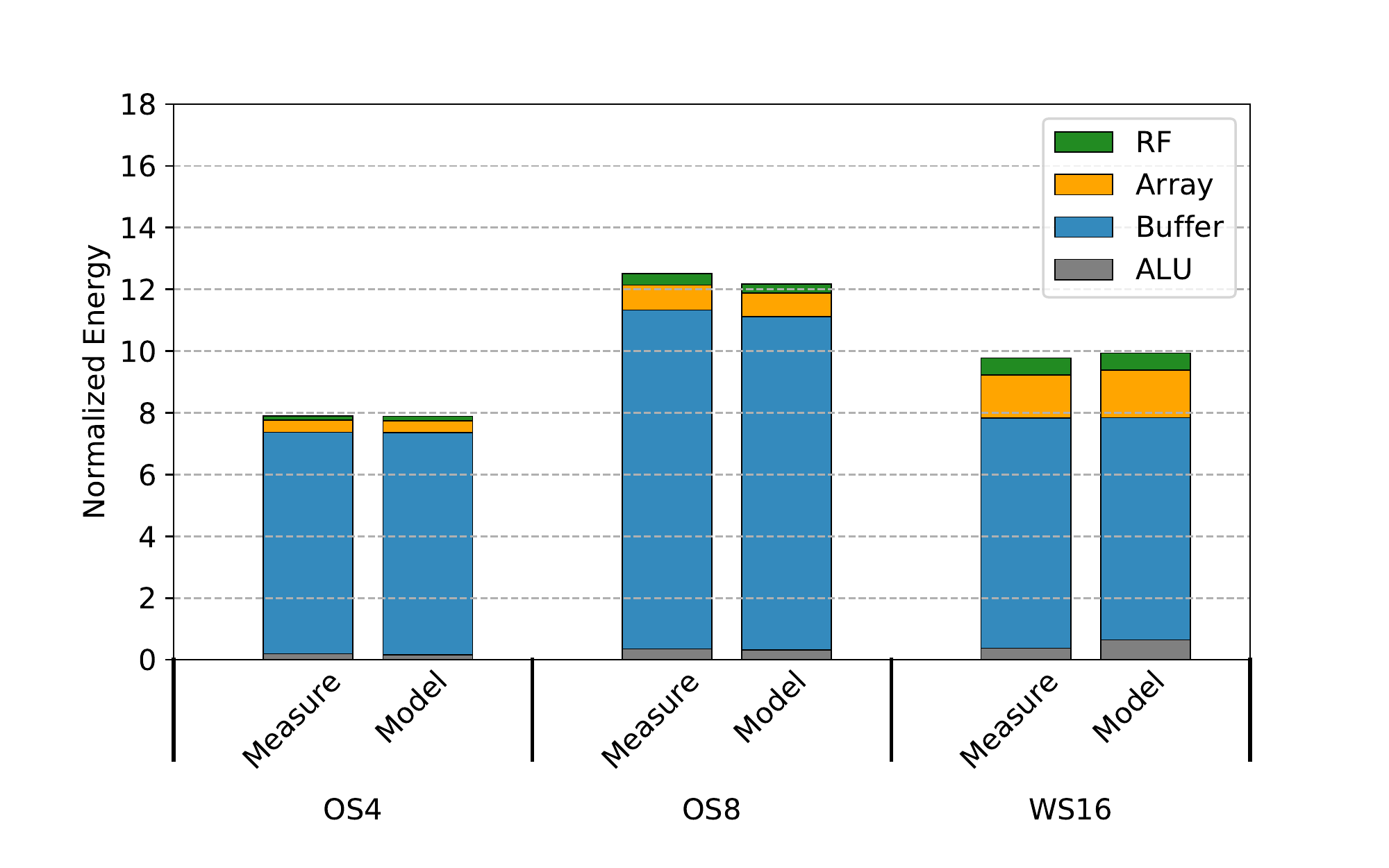} 
        \label{fig:validate_synth}
    }
    \hspace{1em}
    \subfloat[Energy breakdown comparison between reported Eyeriss model and our model.]{
        \centering
        \includegraphics[width=0.45\textwidth]{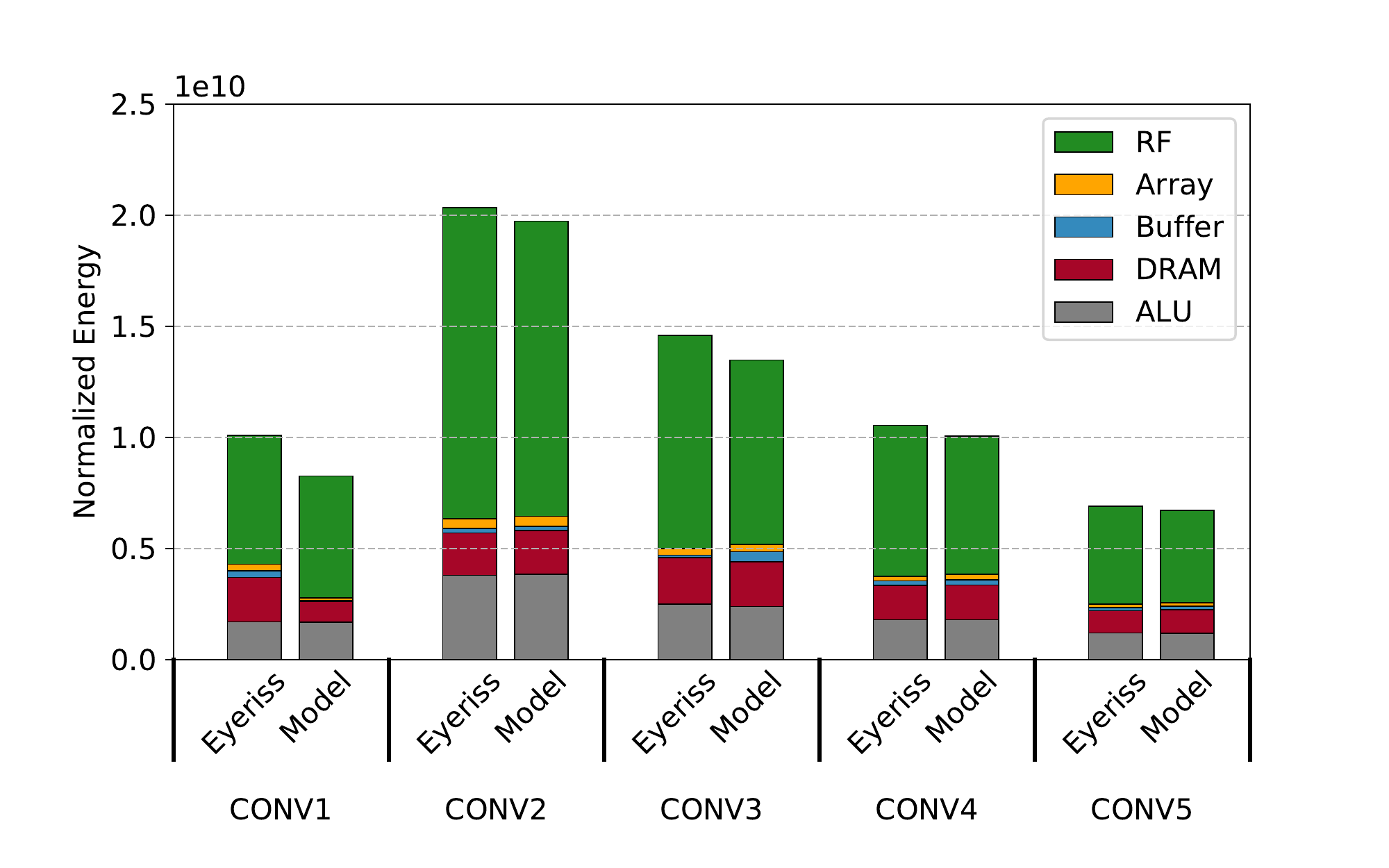} 
        \label{fig:validate_eyeriss}
    }
    \caption{Validation of the analytical model against post-synthesis results and previous published model~\cite{eyeriss_isca16}.}
    \label{fig:model_validate}
\end{figure*}

\textbf{Analysis framework:}
To allow for rapid design exploration, we also developed an analytical model to estimate the performance and energy efficiency of the ASIC DNN accelerators. We use CACTI~6.5~\cite{cacti_micro07} to model SRAM arrays and tune its parameters to match our 28\,nm commercial memory library. For small arrays and register files (RFs), we use the Cadence XtensaProcessor Generator~\cite{tensilica} to extract energy numbers based on our standard cell library. Table~\ref{tab:table_energy} shows the energy cost of accessing memories with different sizes. Note that our energy ratios between memories and MAC are larger than those reported in Eyeriss~\cite{eyeriss_isca16}. There are several reasons: we use a 28\,nm technology instead of 65\,nm; our memory is highly banked with higher energy cost; and our MAC units consume lower energy as their activity factors are relatively low with data stationary patterns. Nevertheless, our analytical framework works with different technology processes, and it is easy to supply new cost models to study more advanced technologies. Also, many of our observations in Section~\ref{sec:result} are technology-independent.

\begin{table}
    \centering
    \caption{Energy per 16-bit access with various register file (RF) and SRAM sizes, and for a MAC operation, one hop communication cost and a DRAM access.}  
    \label{tab:table_energy}
    \begin{threeparttable}
    \begin{tabular}{cc}
        \toprule
        RF Size & Energy (pJ) \\
        \midrule
        16 B  & 0.03 \\
        32 B  & 0.06 \\
        64 B  & 0.12 \\
        128 B & 0.24 \\
        256 B & 0.48 \\
        512 B & 0.96 \\
        \midrule
        MAC & 0.075 \\ 
        Hop & 0.035 \\
        \bottomrule
    \end{tabular}
    \end{threeparttable}
    \quad
    \begin{threeparttable}
    \begin{tabular}{cc}
        \toprule
        SRAM Size & Energy (pJ) \\
        \midrule
        32 KB  & 6 \\
        64 KB  & 9 \\
        128 KB & 13.5 \\
        256 KB & 20.25 \\
        512 KB & 30.375 \\
        \midrule
        DRAM & 200 \\
        \bottomrule
    \end{tabular}
    \end{threeparttable}
\end{table}

To compute the overall memory energy in an $L$-level hierarchy, we adopt a model similar to \cite{eyeriss_isca16} and \cite{timeloop}:
\begin{equation*}
\small
E = \sum_{i=1}^{L}{\mathrm{\#acc}_i \times e_i}
\quad\mathrm{where}\;\;
\mathrm{\#acc}_i = \prod_{j=i}^{L} \mathrm{RT}_j
\end{equation*}
Here $e_i$ is the energy of accessing the $i$th level once. The total numbers of accesses are affected by data reuses $\mathrm{RT}_i$ at different memory levels. It is defined as the number of times the data are accessed by its immediate lower-cost (child) level during its lifetime in this level. To support direct inter-PE communication in systolic arrays, we treat neighbor PEs as an additional level in the hierarchy. We distinguish the cost for different communication distances (Figure~\ref{fig:replication}) which is an improvement over \cite{eyeriss_isca16}.

Since all designs today are power constrained, finding the optimal accelerator design now becomes an optimization problem of minimizing $E$ over the 3D design space, similar to Yang's work~\cite{Xuanthesis}. $e_i$ is determined by the resource allocation (Table~\ref{tab:table_energy}), and $\mathrm{RT}_i$ can be directly calculated from the dataflow and loop blocking schemes. In this work, we simply perform a conservatively pruned search over the full design space guided by domain-specific knowledge. This analytical model is available at \url{https://github.com/xuanyoya/Interstellar-CNN-scheduler}.

\begin{table}
\centering
\caption{ASIC designs for model validation.}  
\label{tab:validate_designs}
\begin{threeparttable}
\begin{tabular}{cccccc}
    \toprule
    Name & Dataflow & PE Array & RF & SRAM \\
    \midrule
    OS4 & $X$ & 1D, 4 & 32\,B & 32\,KB \\
    OS8 & $X$ & 1D, 8 & 64\,B & 64\,KB \\
    WS16 & $C\!\mid\!K$ & 2D, 4$\times$4 & 64\,B & 32\,KB \\
    \bottomrule
\end{tabular}
\end{threeparttable}
\end{table}

\textbf{Framework validation:}
We have thoroughly validated the accuracy of our model by comparing its results to complete designs generated by our synthesis toolchain. Table~\ref{tab:validate_designs} shows three example designs we have generated in ASIC platforms, and Figure~\ref{fig:model_validate} shows the energy comparison between our analytic model and post-synthesis results. The resulting errors are less than 2\%. Furthermore, our framework is also able to reproduce the results from \cite{eyeriss_isca16} with small differences.

%% file: result.tex
\section{Results}
\label{sec:result}

Using our dataflow taxonomy and the ability to rapidly generate and evaluate large numbers of accelerator designs with Halide, we first explore different dataflow and loop blocking choices, and then consider hardware resource optimizations. At the end we leverage the characteristics of these results to introduce an efficient optimizer for DNN accelerators.

\subsection{Impact of Dataflow and Loop Blocking}
\label{sec:res_dataflow}

\begin{figure}
    \centering
    \subfloat[Batch 16 (AlexNet).]{
        \centering
        \includegraphics[width=0.235\textwidth]{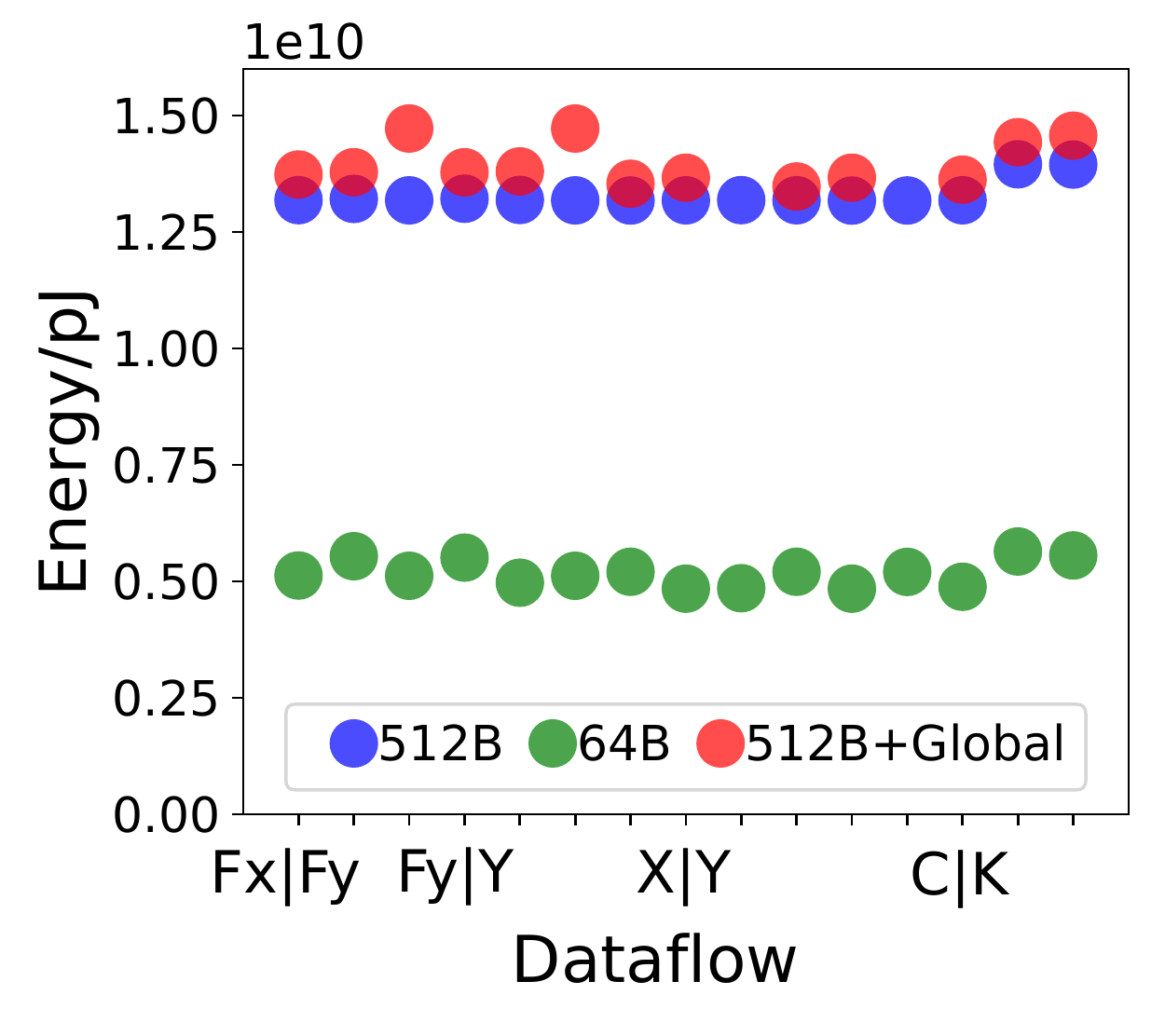}
        \label{fig:df_space:efficiency}
    }
    \subfloat[Batch 1 (AlexNet).]{
        \centering
        \includegraphics[width=0.235\textwidth]{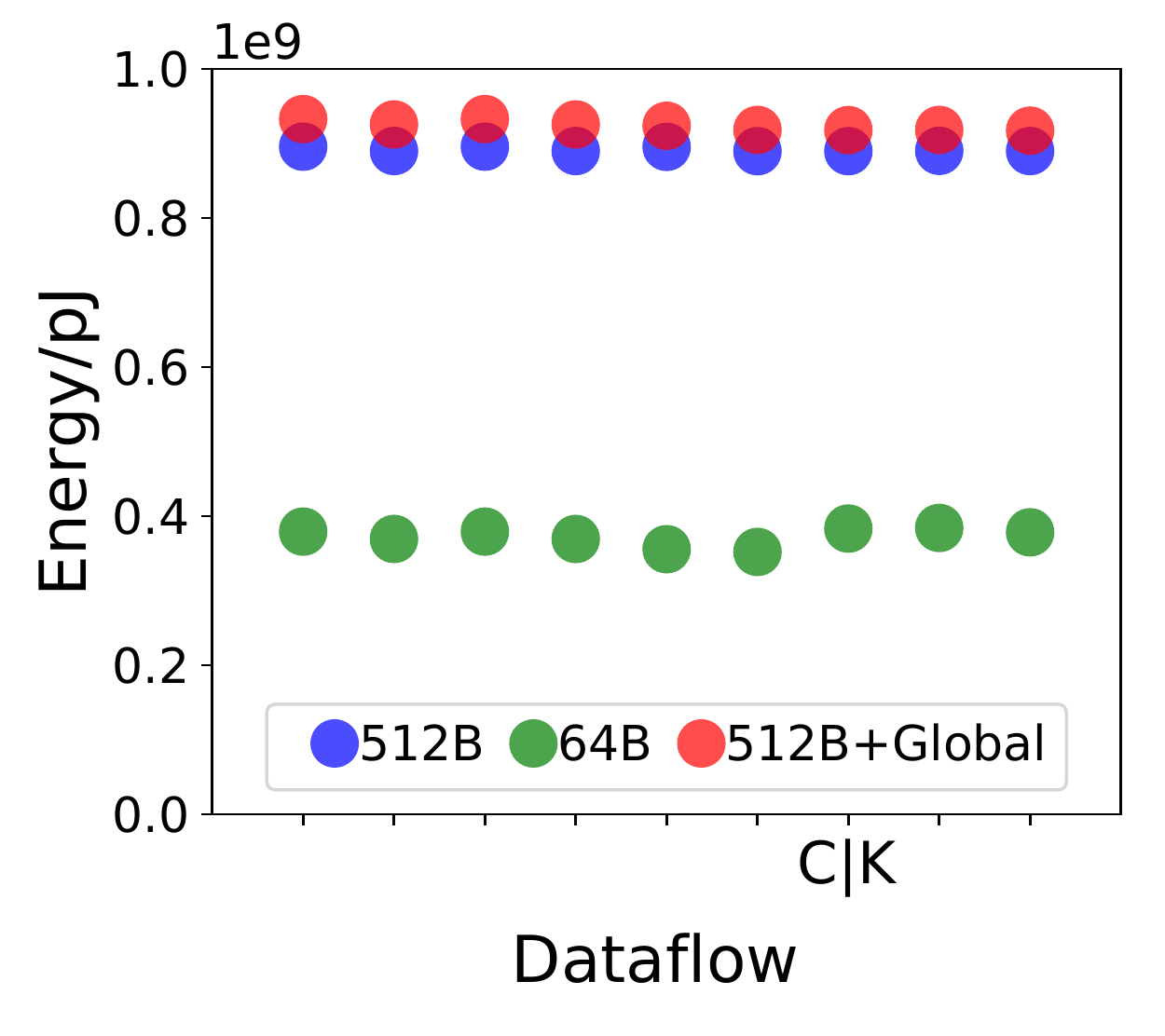}
        \label{fig:df_space:efficiency_batch1}
    }\hspace{-1em}
    \subfloat[Batch 16 (GoogleNet).]{
        \centering
        \includegraphics[width=0.235\textwidth]{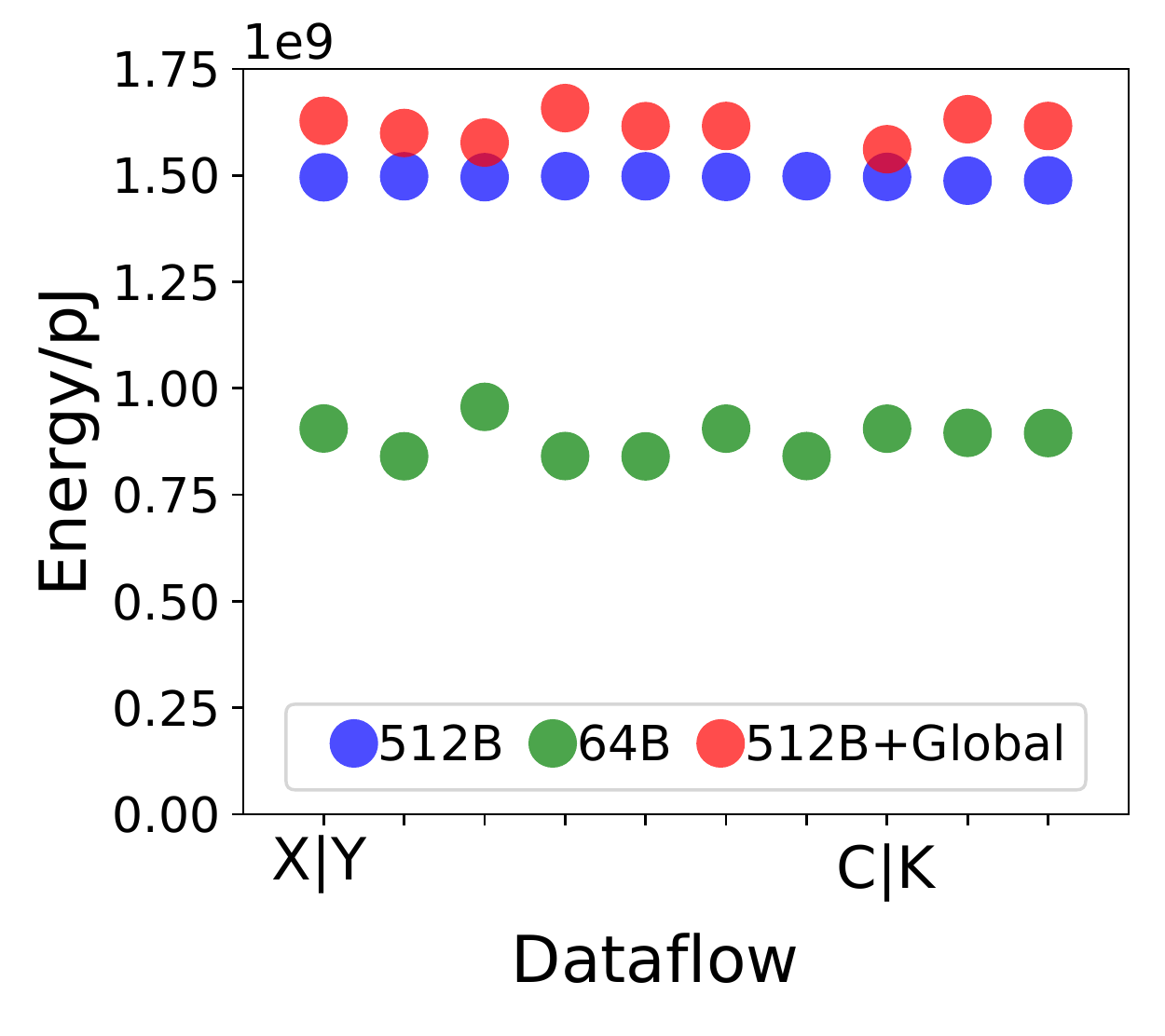}
        \label{fig:df_space:efficiency_googlenet}
    }
    \subfloat[Batch 1 (GoogleNet).]{
        \centering
        \includegraphics[width=0.235\textwidth]{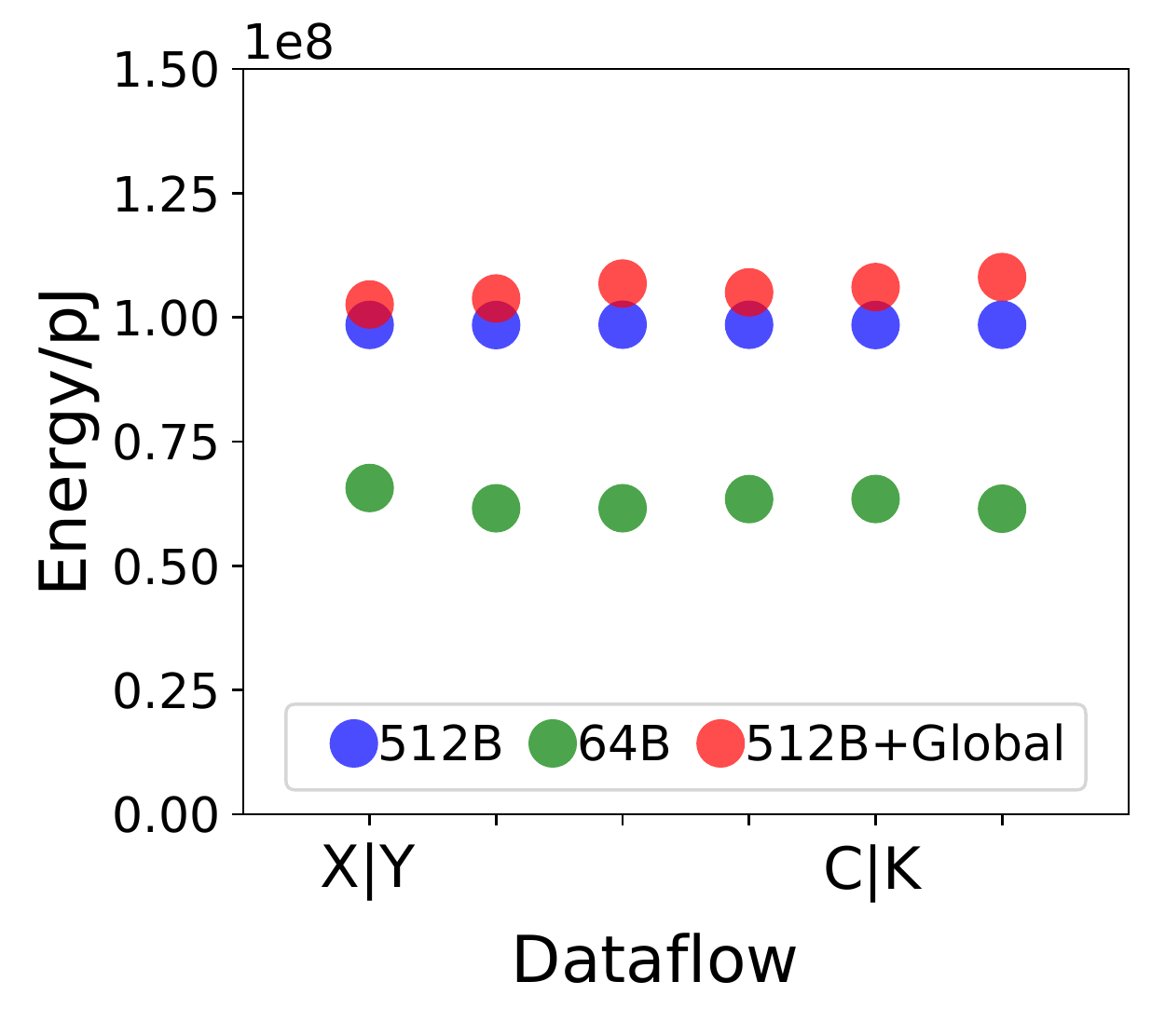}
        \label{fig:df_space:efficiency_batch1_googlenet}
    }    \hspace{-1em}
    \caption{Design space of dataflow for AlexNet CONV3 and GoogLeNet 4C3R layers. The Y-axis is the energy consumed to execute the entire batch. Different dataflows are shown horizontally, with only the most common choices labeled for clarity. All dataflows use replication and the optimal loop blocking schemes. Different colors represent different hardware resource allocations.}
    \label{fig:df_space}
\end{figure}


Figure~\ref{fig:df_space} compares the energy efficiency of different dataflow choices. We use the CONV3 layer in AlexNet and 1$\times$1 reduction layer 4C3R in GoogLeNet Inception (4c) module as examples. The other layers have also been investigated, and share a similar trend. More DNNs will be studied in Section~\ref{sec:auto-optimizer}. We use the optimal loop blocking scheme for each dataflow. We can see that when optimized loop blocking schemes are applied, many different dataflows achieve similar and close-to-optimal energy efficiency on the same hardware configuration. We have evaluated three different hardware configurations: the blue one is the same as Eyeriss~\cite{eyeriss_isca16}, with 512\,B register file (RF), 128\,KB SRAM buffer, and 16$\times$16 PE array per PE; the red one uses a different array bus design, which disables inter-PE communication and broadcasts all data from the global buffer; the green one uses a smaller 64\,B RF to lower its access energy (see Section~\ref{sec:res_memory}). Figure \ref{fig:df_space:efficiency_batch1} and \ref{fig:df_space:efficiency_batch1_googlenet} also show the cases with batch size 1, which is most commonly used in mobile systems; the conclusion is consistent across different batch sizes. The small influence of dataflow remains true over a wide set of experiments including different layer types, different PE array structures and sizes, different memory configurations, and different energy cost models.

\begin{figure}[tb]
    \centering
    \subfloat[No Replication (AlexNet)]{
        \centering
        \includegraphics[width=0.23\textwidth]{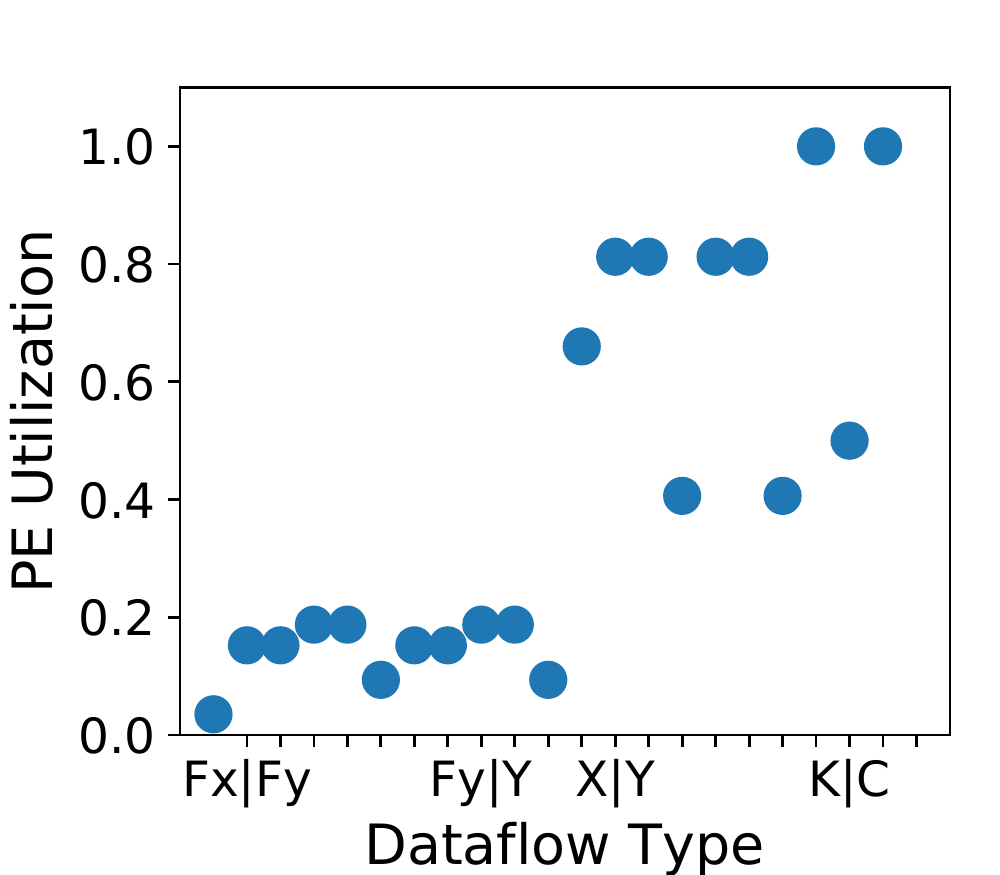}
        \label{fig:utilization:no_rp}
    }\\
    \subfloat[With Replication (AlexNet)]{
        \centering
        \includegraphics[width=0.23\textwidth]{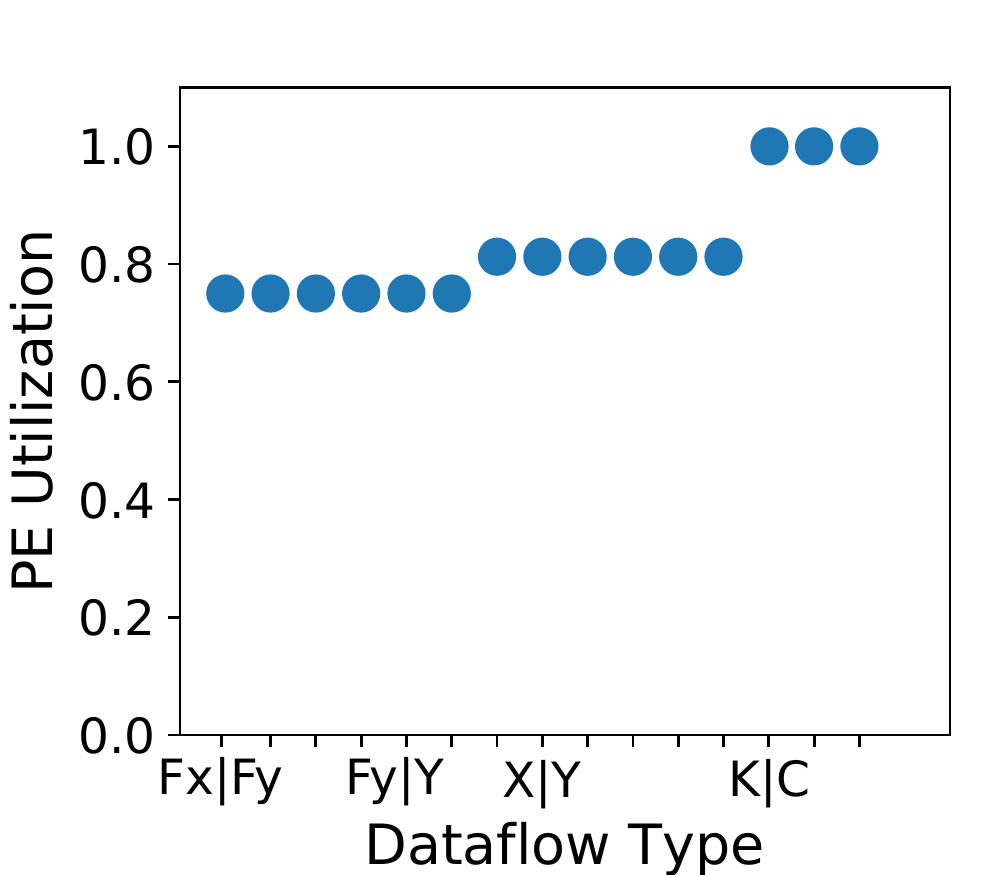}
        \label{fig:utilization:with_rp}
    }
    \subfloat[With Replication (GoogleNet)]{
        \centering
        \includegraphics[width=0.23\textwidth]{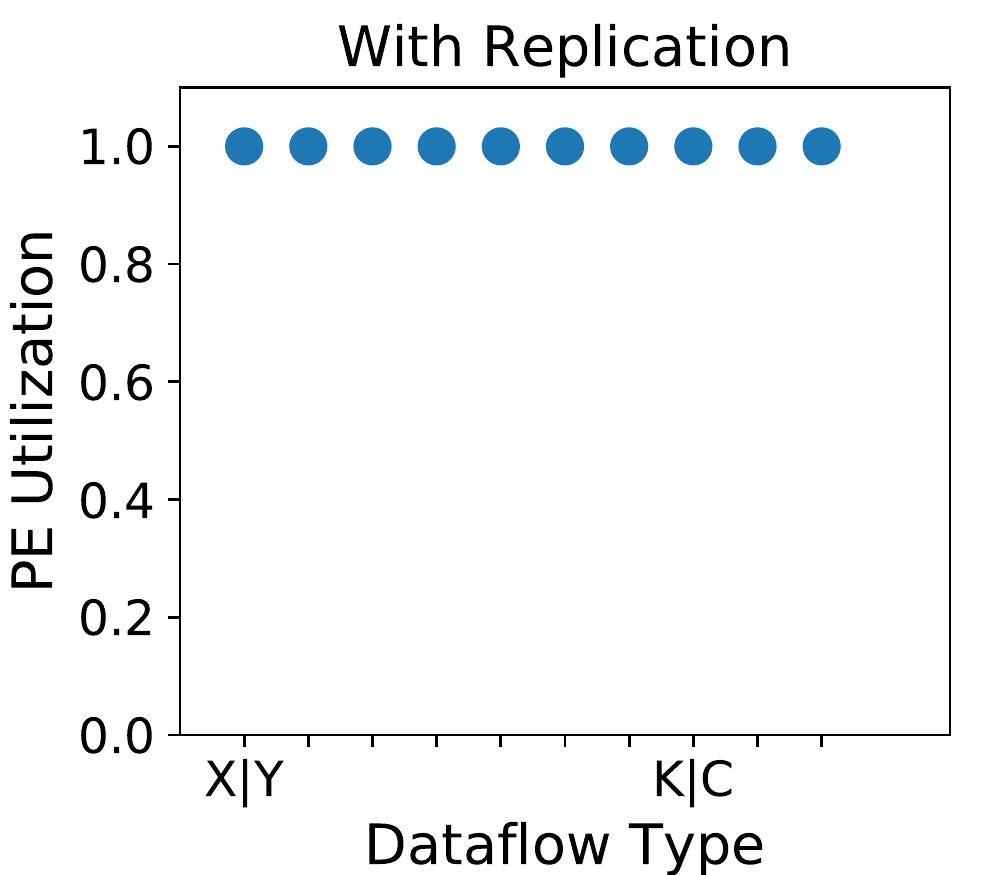}
        \label{fig:utilization:with_rp_googlenet}
    }
    \caption{PE array utilization for the energy-optimal dataflow choices on AlexNet CONV3 layer with and without replication, and GoogleNet 4C3R layer with replication.}
    \label{fig:utilization}
\end{figure}

On the other hand, in Figure~\ref{fig:utilization} we observe that the PE array utilization (active PE ratio for each run), and therefore the computation throughput, is more sensitive to different dataflow choices than the energy efficiency for some convolution layers. Without replication (Figure~\ref{fig:utilization:no_rp}), the overall utilization can vary significantly and stay low for many dataflow choices. However, using proper replication substantially improves the utilization and eliminates most of the differences among all dataflows (Figure~\ref{fig:utilization:with_rp} and \ref{fig:utilization:with_rp_googlenet}). These results also imply that accelerators that support a diversity of replication schemes, such as CGRAs and Eyeriss V2~\cite{chen2018eyeriss}, will generally achieve higher overall utilization compared to the ones with fixed interconnects. Here, the impact of the interconnect bandwidth to evaluate the overall performance is not included in the analysis, as prior works~\cite{chen2018eyeriss, DBLP:journals/corr/abs-1805-02566} have already studied such impact.  Figure~\ref{fig:utilization:with_rp} also demonstrates, for the CONV3 layer in AlexNet, that the $C\!\mid\!K$ dataflow achieves 20\% higher utilization than the others such as $F_Y\!\mid\!Y$. This is because the channel dimensions $C$ and $K$ are typically the largest in most CONV and FC layers, so it is easier to unroll them onto a fixed-sized PE array with small fragmentation. For this reason, we will use the $C\!\mid\!K$ dataflow in the rest of this paper.

Not all the computational layers have as much data sharing to exploit.  Weight sharing in FC layers only comes from batching, and some applications limit the batch size to be small, even one. Interestingly, even in these computations the dataflow does not have a large influence on performance or energy. For computations with limited reuse, the data must come from the off-chip DRAM, or the last level on-die storage, if it is large enough.  The storage properties at this level will limit the device's energy and performance, so for this class of application, the design of the computation units is less important. 


\begin{figure}
    \centering
    \includegraphics[width=0.45\textwidth]{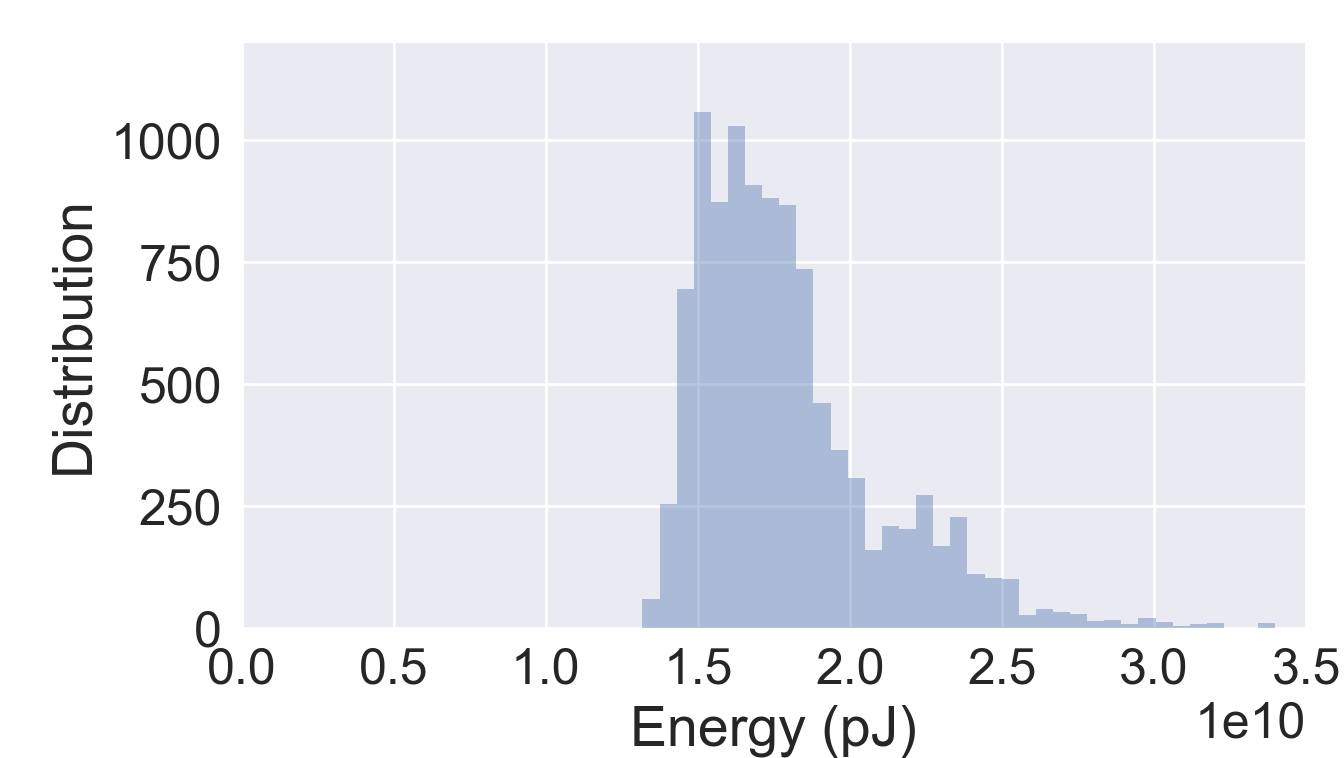}
    \caption{Design space of loop blocking for AlexNet CONV3 using dataflow $C\!\mid\!K$ with 512\,B RF per PE.}
    \label{fig:blocking_dist}
\end{figure}

Instead of dataflow choices, Figure~\ref{fig:blocking_dist} shows the design space of loop blocking for AlexNet CONV3, using a 512\,B RF, corresponding to the blue configuration in Figure~\ref{fig:df_space:efficiency}. The energy variance of different blocking schemes is much more significant than that of dataflow, and only 30\% of the schemes fall within 1.25$\times$ of the minimum energy. This indicates that loop blocking has a large impact on energy efficiency.

\textbf{Observation 1:}
\emph{With the same hardware resources, many different dataflows are able to achieve similar and close-to-optimal energy efficiency, as long as proper loop blocking and replication are used.}

\begin{figure}
    \centering
    \includegraphics[width=0.5\textwidth]{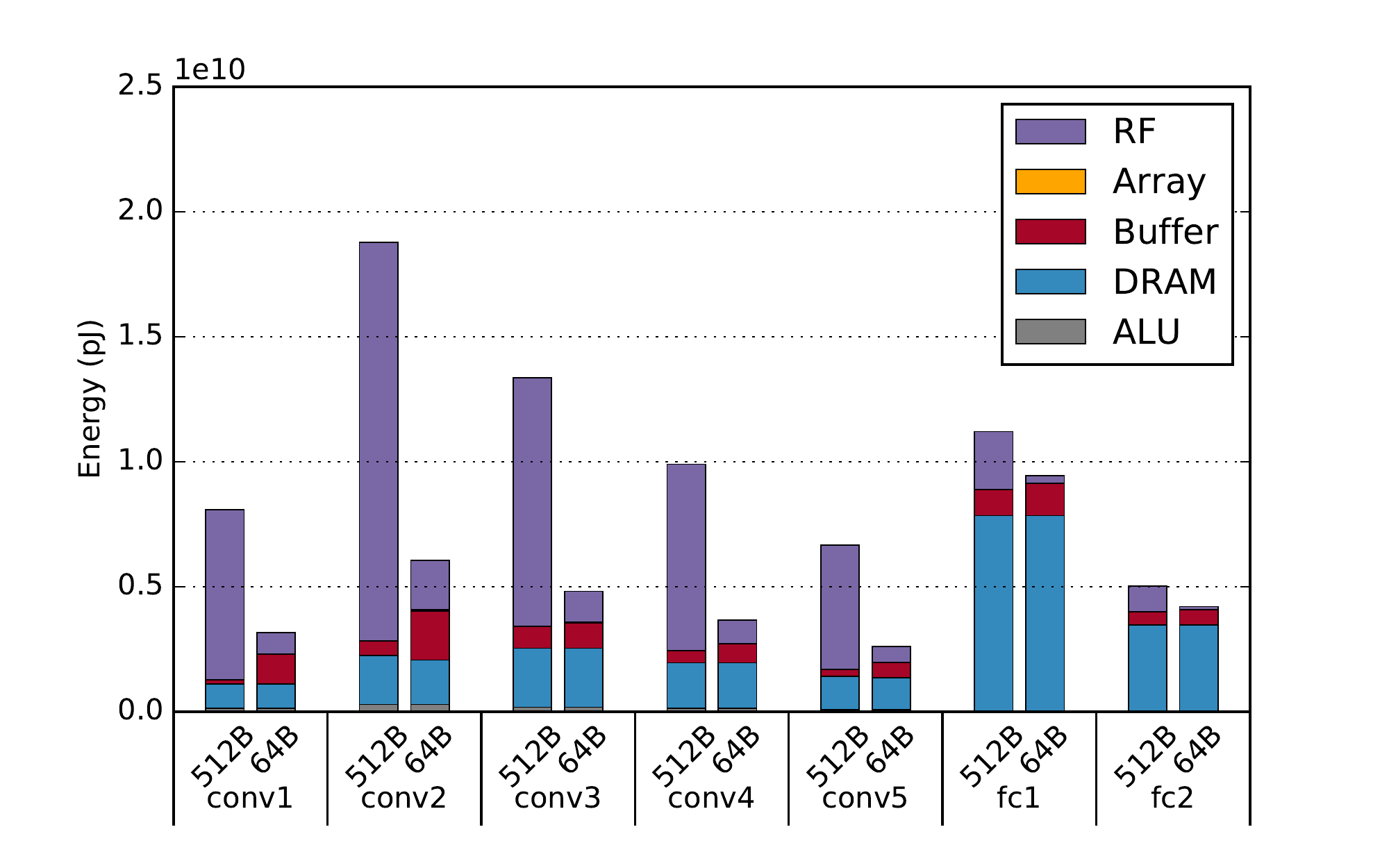}
    \caption{Energy breakdown comparison between 512\,B and 64\,B RF sizes with the same dataflow. Using a 64\,B RF reduces the overall energy significantly.}
    \label{fig:mem_breakdown}
\end{figure}

In hindsight, this result is not surprising. When the DNNs exhibit enormous data reuse opportunities, as long as high data reuse is achieved through proper loop blocking schemes, the resulting energy efficiency should be good. When the reuse is limited, the performance is limited by the bandwidth of the last level in the memory hierarchy instead of the PE array. These two situations are further illustrated in Figure~\ref{fig:mem_breakdown}, where the left bars with 512\,B RF show the energy breakdown of the optimal dataflow for the blue configuration in Figure~\ref{fig:df_space:efficiency}. For CONV layers with high reuse, most energy is consumed in the RF level rather than the array buses or intermediate buffers. By optimally blocking the computation, nearly all accesses (98\%) occur at the RF level, making it the dominant energy component. For FC layers with limited reuse, most of the DRAM energy is inevitable, since the data have to be fetched at least once from off-chip (compulsory misses). On the other hand, the on-chip communication is generally only a small portion of the total energy, and therefore blocking choice has a more substantial impact on the overall energy efficiency than dataflow choice.

\subsection{Impact of Hardware Resource Allocation}
\label{sec:res_memory}

\begin{figure}[tb]
    \centering
    \includegraphics[width=0.45\textwidth]{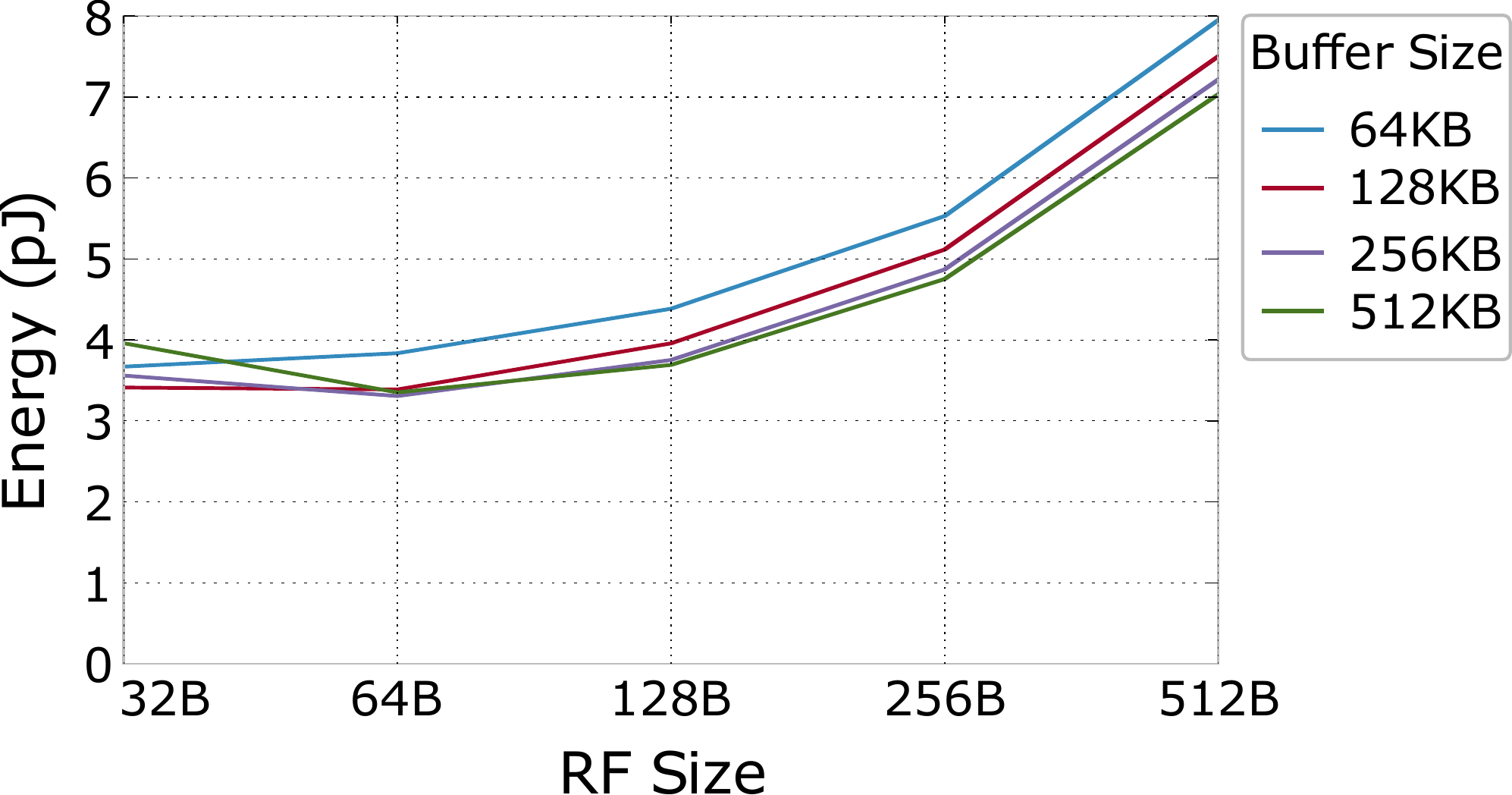}
    \caption{Memory hierarchy exploration with dataflow $C\!\mid\!K$. Different RF sizes per PE are shown horizontally. Lines with different colors correspond to different SRAM buffer sizes.}
    \label{fig:mem_search}
\end{figure}

Another interesting result from Figure~\ref{fig:mem_breakdown} is that the total energy is always dominated by the RF level with a 512\,B RF. This result indicates that this resource allocation may be suboptimal. Figure~\ref{fig:mem_search} shows the impact of memory resource allocation on energy efficiency. The energy is accumulated across all layers (including FC layers) in AlexNet, and contains both computation and memory access portions. It indicates that using a smaller RF size such as 32 or 64\,B can improve the total energy efficiency by up to 2.6$\times$. If we also increase the global SRAM buffer size, the energy efficiency can further improve. However, when SRAM buffer size grows beyond 256\,KB, the benefit becomes negligible. Given the significant area cost, it is not always necessary to use large global buffers.

The right bars in Figure~\ref{fig:mem_breakdown}, which give the energy breakdown of using a 64\,B RF, illustrate that the energy decreases dramatically for all the CONV layers due to the much lower energy cost per access of the smaller RF. At the same time, more accesses go to the inter-PE array level and the global buffer, since the smaller RF captures less data reuse inside each PE. But reducing the RF size has almost no impact on the DRAM energy, as the data are still efficiently reused in the global buffer. Overall, a smaller RF achieves significantly better energy efficiency, with a more balanced energy breakdown among different memory hierarchy levels.

\begin{figure}[tb]
    \centering
    \includegraphics[width=0.45\textwidth]{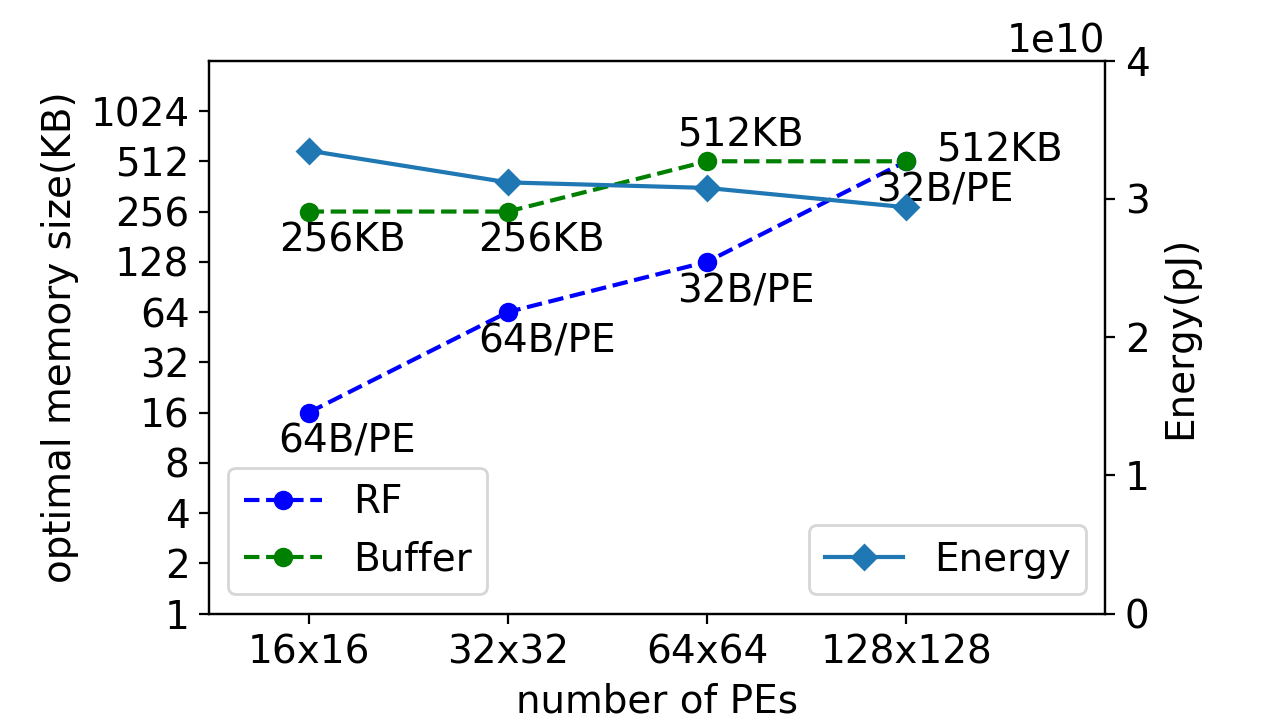}
    \caption{The optimal memory resource allocation and the corresponding total energy when varying PE array size.}
    \label{fig:vary_pe}
\end{figure}

\begin{figure*}[htbp]
    \centering
    \includegraphics[width=\textwidth]{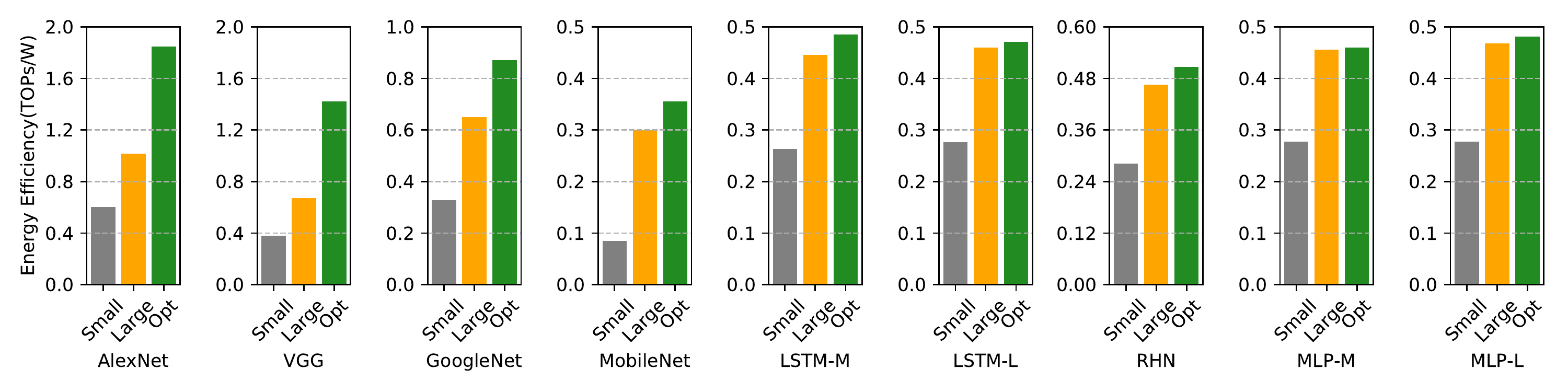} 
    \caption{Overall energy efficiency improvement by using the auto-optimizer.}
    \label{fig:optimizer_overall}
\end{figure*}

\textbf{Observation 2:}
\emph{The total energy of an efficient system should not be dominated by any individual level in the memory hierarchy.}

Observation 2 also explains why some output-stationary and weight-stationary designs do not perform well, as discussed by \cite{eyeriss_isca16}. Those designs cannot capture sufficient reuse at the RF level, and result in high energy consumption at the DRAM level, which dominates the overall energy.

However, there is an exception for Observation 2. When DRAM dominates the total energy but the number of DRAM accesses is already minimized (fetching the input once and writing back output once), the DNN is memory bound, and based on Amdahl's law, little further optimization can be achieved for the memory hierarchy. This is the case particularly for a batch size of 1, and MLPs and LSTMs that contain many FC layers.


We also investigate whether changing the hierarchy itself can further improve the energy efficiency. Due to the dominant role of the RF level, we add another level of private register file.
The largest energy efficiency improvement is obtained when sizing each memory level is based on the rule that the ratio of the on-chip storage sizes between the adjacent levels should be around 4 to 16. Specifically, the overall energy efficiency of AlexNet is improved by 25\%, by choosing 16\,B and 256\,B to be the two level register file sizes, and 256\,KB to be the global buffer size.

Figure~\ref{fig:vary_pe} depicts the optimal memory resource allocation and the corresponding total energy for AlexNet when varying PE array size. We use only one level of RF here. These correspond to the optimal points on the optimizing plane shown in Figure~\ref{fig:3d_space}. With increasing numbers of PEs, the optimal memory size at each level grows sub-linearly. Ideally we would like to keep the same amount of data reuse with constant storage capacity for each PE, which would lead to linearly increased memory size. However, the access cost of each memory level grows with its size (Table~\ref{tab:table_energy}), which slows down the optimal capacity scaling to sub-linear. Between the RF and the SRAM buffer, the data reuse in RF is more critical. So the RF level has a stronger trend to keep constant capacity per PE. But it is eventually bounded by the size of the next-level, i.e., the SRAM buffer.

Also we notice that the total energy reduces slightly with the increasing number of PEs. This indicates that larger arrays will have a significant effect on throughput and a small change in energy efficiency. The energy improvement is achieved by buffering more data on chip for reuse, and since most communication is nearest neighbor, the larger die does not increase communication costs significantly. 




\subsection{An Efficient Optimizer}
\label{sec:auto-optimizer}

With the large number of hardware and software choices for DNN accelerators, exhaustive search for the optimal designs is usually infeasible. Instead, using the observations above, we can speed up the optimization process by pruning the search space and evaluating only a small number of candidates using the framework from Section~\ref{sec:method}.

We developed an auto-optimizer that efficiently finds energy efficient accelerator designs for given DNNs. The optimizer takes as input the DNN topology, the energy cost model, and various constraints such as the total chip area. First, according to \textbf{Observation 1}, we fix the dataflow to be $C\!\mid\!K$, and only search the design points on the optimizing plane in Figure~\ref{fig:3d_space}. Next, we only evaluate a subset of hardware configurations with the optimal size of each memory level satisfying \textbf{Observation 2}, leveraging the rule that the ratio of the on-chip storage sizes and the adjacent levels should be around 4 to 16. It outputs an optimized design with corresponding Halide schedule primitives, which can then be fed into our hardware synthesis toolchain.

We use four CNNs, three LSTMs, and two MLPs as benchmarks to demonstrate the effectiveness of our efficient optimizer. All DNNs evaluated use 16-bit precision. The CNNs are AlexNet, VGG-16~\cite{vggnet_arxiv14}, MobileNet~\cite{DBLP:journals/corr/HowardZCKWWAA17}, and GoogleNet~\cite{googlenet_arxiv14} with batch size 16. The LSTM-M and LSTM-L are proposed by Google for sequence-to-sequence learning~\cite{DBLP:journals/corr/SutskeverVL14} with embedding sizes 500 and 1000. We also study the Recurrent Highway Network (RHN)~\cite{DBLP:journals/corr/ZillySKS16}. The MLPs are from \cite{prime_isca16} with batch size 128. 
We use two baselines, both using dataflow $C\!\mid\!K$, which are the two left columns in Figure~\ref{fig:optimizer_overall}. The smaller chip uses a memory hierarchy similar to Eyeriss~\cite{eyeriss_isca16}, and 16$\times$16 PE array, whose area and power budgets are suitable for mobile platforms. The larger chip uses 128$\times$128 PE array with a 8\,B register per PE, 64\,KB for the first-level global buffer, and a 28 MB second-level global buffer, similar to cloud-based accelerators such as TPU~\cite{tpu_isca17}.


Figure~\ref{fig:optimizer_overall} demonstrates the energy efficiency gain achieved by the efficient optimizer. We can improve the energy efficiency by up to 3.5$\times$, 2.7$\times$, and 4.2$\times$ for VGG-16, GoogleNet and MobileNet, up to 1.6$\times$ for LSTMs, and up to 1.8$\times$ for MLPs. The optimal memory hierarchy uses 16\,B and 128\,B for the first-level and second-level register files, with a 256\,KB global SRAM double buffer. This hardware configuration is shared by all the layers in the DNNs. The overall system energy consumption is not dominated by the RF level. The energy efficiency for the nine benchmarks are 1.85, 1.42, 0.87, 0.35, 0.49, 0.47, 0.5, 0.46, and 0.48 TOPs/W, respectively. Notice that even though the larger system has a smaller RF size, its energy is better than the smaller system. This is because with a much larger global SRAM buffer, it can store all the input and output data and the layer weights, and the accesses to DRAM are eliminated when switching to the next layer.

%% file: conclusion.tex
\section{Conclusion}
\label{sec:conclusion}


To help elucidate factors that matter for DNN accelerators, we realized that Halide's scheduling language was almost rich enough to describe the design space of all possible accelerators. By extending the scheduling language to include local communication, and creating a hardware backend, we were able to generate and hence fairly compare all proposed dense DNN accelerators.  The results show that as long as the data reuse and the resource utilization are maximized by proper loop blocking and mapping replication, many hardware dataflow choices will be near optimal. Not surprisingly, optimizing the memory hierarchy had a large influence on energy efficiency, with smaller local registerfiles and deeper memory hierarchies providing the best performance. 

%% file: ack.tex
\section{Acknowledgments}

The authors want to thank Kayvon Fatahalian, Jonathan Ragan-Kelley, Andrew Adams and Stephen Richardson for the insightful
discussion, Keyi Zhang for his help on LaTex formatting, and the anonymous reviewers for their valuable comments. This work was supported by the DSSoC DARPA grant, the Stanford AHA Agile Hardware Center and Affiliates Program, the Stanford SystemX Alliance, the Stanford Platform Lab, and SRC Center for Research on Intelligent Storage and Processing-in-memory
(CRISP).